\documentclass[12pt,preprint]{aastex}

\shorttitle{Spitzer observations of polars}
\shortauthors{Brinkworth et al.}

\begin{document}

\title{Spitzer Space Telescope observations of magnetic cataclysmic 
variables:\ possibilities for the presence of dust in polars}

\author{C.~S. Brinkworth, D.~W. Hoard, S. Wachter}
\affil{Spitzer Science Center, California Institute of Technology, 
MS 220-6, 1200 E.\ California Blvd., Pasadena, CA 91125}

\author{S.~B. Howell}
\affil{WIYN Observatory/National Optical Astronomy Observatory, 
950 N.\ Cherry Ave., Tucson, AZ 85719}

\author{David R. Ciardi}
\affil{Michelson Science Center, California Institute of Technology, 
MS100-22, 770 S. Wilson Ave., Pasadena, CA 91125}

\author{P. Szkody}
\affil{Department of Astronomy, University of Washington, Box 351580, 
Seattle, WA 98195-1580}

\author{T.~E. Harrison}
\affil{Department of Astronomy, New Mexico State University, Box 30001, 
MSC 4500, Las Cruces, NM 88003}

\author{G.~T. van Belle}
\affil{Michelson Science Center, California Institute of Technology, 
MS 100-22, 770 S.\ Wilson Ave., Pasadena, CA 91125}

\author{A.~A. Esin}
\affil{Department of Physics, Harvey Mudd College, 301 E.\ 12th St., 
Claremont, CA 91711-5990}

\begin{abstract}
We present Spitzer Space Telescope photometry of six short-period polars, 
EF Eri, V347 Pav, VV Pup, V834 Cen, GG Leo, and MR Ser. We have combined 
the Spitzer Infrared Array Camera (3.6--8.0 $\mu$m) data with the 2MASS 
$JHK_{\rm s}$ photometry to construct the spectral energy distributions 
of these systems from the near- to mid-IR (1.235--8 $\mu$m). We find 
that five out of the six polars have flux densities in the 
mid-IR that are substantially in excess of the values expected from the 
stellar components alone. We have modeled the observed SEDs with 
a combination of contributions from the white dwarf, secondary star, 
and either cyclotron emission or a cool, circumbinary dust 
disk to fill in the long-wavelength excess. We find that a circumbinary 
dust disk is the most likely cause of the 8$\mu$m excess in all cases, 
but we have been unable to rule out the specific (but unlikely) 
case of completely 
optically thin cyclotron emission as the source of the observed 8$\mu$m 
flux density. While both model components can generate enough flux at 8$\mu$m, 
neither dust nor cyclotron emission alone can match the excess above 
the stellar components at all wavelengths. A model combining both 
cyclotron and dust contributions, possibly with some accretion-generated 
flux in the near-IR, is probably required, but our observed SEDs are not 
sufficiently well-sampled to constrain such a complicated model. 
If the 8 $\mu$m flux density \emph{is} caused by the 
presence of a circumbinary dust disk, then 
our estimates of the masses of these disks are many 
orders of magnitude below the mass required to affect CV evolution. 
\end{abstract}


\keywords{cataclysmic variables (V834 Cen, EF Eri, GG Leo, V347 Pav, 
VV Pup, MR Ser) --- stars: magnetic fields --- stars: low-mass --- 
infrared: stars }

\section{Introduction}

Cataclysmic variables (CVs) are semi-detached binary systems
consisting of a white dwarf (WD) primary star that accretes material
from a low mass secondary star, with typical orbital periods of 
$P_{\rm orb} \lesssim 1$ day. 
Due to its large specific orbital angular momentum, the mass lost from 
the secondary star doesn't fall directly onto the WD. 
In most CVs, it instead settles into a disk around the primary star
before losing enough angular momentum through viscous processes to
finally accrete onto the WD. 
The release of gravitational potential energy in the disk causes it 
to be, by far, the brightest component of a CV over a wide range of 
wavelengths. 
On the other hand, polars (or AM Her stars, after the prototype) are 
CVs containing a WD with a strong ($B \geq 10$ MG) magnetic field. 
The matter lost from the secondary star is captured by the field lines 
and funnelled onto the magnetic pole(s) of the WD. 
The infalling matter is shocked as it nears the surface of the WD, 
radiating over most of the electromagnetic spectrum from X-rays to 
infrared. 
Observational characteristics and types of CVs are extensively reviewed 
by \citet{warner95}.

Polars are highly variable, both at a low level (usually $\sim$1--2 mag) 
over their orbital periods as the geometry of the system changes along 
the line of sight, and on much longer timescales with much greater 
changes in luminosity (up to $\sim$4 mag). 
Until recently, polars were generally thought to hover around a maximum 
brightness, occasionally switching into relatively short low states, 
with no preferred minimum brightness for a given system. 
The cause of these low states is not fully understood, but is possibly 
due to the presence of star spots on the surface of the secondary star 
passing across the L1 point and suppressing mass transfer into the
Roche lobe of the WD \citep{hessman00}.  
The resultant lowered accretion rate leads to a period of lower 
luminosity until the star spot has moved away from the L1 point and 
normal mass transfer resumes. 
More recently, however, a number of polars have been seen to hover in 
low states (\citealt{gerke06}; \citealt{araujo05}), suggesting that the 
true cause of the accretion state changes in these CVs is more 
complex than previously believed. 

The secular evolution of CVs is driven by the loss of orbital angular
momentum, which causes the separation of the two component stars (and,
hence, their orbital period) to continually decrease. 
While the secondary star radius decreases with mass-loss, this 
simultaneous draining of angular momentum from the system keeps it 
in contact with its Roche lobe, and enables the ongoing transfer of 
matter to the WD. 
Observationally, this results in CVs with shorter orbital periods 
having secondary stars with correspondingly later spectral types 
(i.e., lower masses). 
Eventually this ongoing mass-loss from the secondary star causes it 
to drop below the hydrogen-burning limit, and the secondary star 
becomes a degenerate brown-dwarf-like object. 
This change in the internal structure of the secondary star causes 
its radius to \emph{increase} in response to further mass-loss. 
This, in turn, results in a reversal of the trend of orbital period 
change, leading to a gradual increase in both $P_{\rm orb}$ and 
binary separation. 
This is believed to be responsible for the observed minimum orbital 
period of CVs at $P_{\rm orb} \sim 80$ min (\citealt{paczynski81}; 
\citealt{rappaport82}; \citealt{howell01b}).

Population synthesis studies predict that the secondary stars in 
$\sim70$\% of all CVs should have evolved past the hydrogen-burning limit 
(e.g., \citealt{kolb93}; \citealt{howell97}).
\citet{littlefair03} reviewed all relevant observational 
data and analyses in the literature and concluded that while 
there is plausible indirect evidence for the existence of brown 
dwarf secondary stars in 39 systems, none of these had a reliable 
secondary star mass estimate or accurate enough spectral type to 
be certain.  
More importantly, they concluded that even if all of the potential 
candidates do indeed contain brown dwarf secondary stars, this only 
amounts to about 10\% of all CVs -- a huge deficit compared to the 
predicted 70\% of CVs.
Consequently, it has become very important to ascertain the number 
of systems, if any, containing brown dwarf secondary stars, in order 
to place limits on our theories of CV population age and evolution.

Here we present Spitzer Space Telescope photometry of six polars that
were identified as having very short orbital periods 
($P_{\rm orb} \lesssim 90$ min) and/or very red 2MASS colors \citep{hoard02}. 
This, along with other published evidence (e.g., compiled by 
\citealt{littlefair03}), suggests that they are good candidates for 
containing brown dwarf secondary stars. 
Our Spitzer observations were intended to enable us to 
unambiguously classify the 
spectral type of the secondary star by isolating its contribution 
to the mid-IR flux.  
However, as shown in \citet[][henceforth, Paper I]{howell06a},
even in the mid-IR the signature of the low mass 
secondary star in these polars is contaminated by excess emission beyond 
the level expected from the stellar components.  In this paper, 
we explore in detail the possible origins of the mid-IR emission in 
magnetic CVs. 

Preliminary results for four of our targets (EF Eridani, GG Leonis, 
V347 Pavonis, and RX J0154.0-5947) were presented in Paper I.
The first three are presented again here with more extensive modeling 
for completeness of our entire sample of polars, but RX J0154.0$-$5947 
is poorly studied and too little is known about its system parameters 
to warrant the more detailed modeling in this paper
(in addition, it is the faintest of our polars -- see Paper I). 
In addition, we present the first results from Spitzer mid-IR 
observations of the polars V834 Centauri, VV Puppis, and MR Serpentis.

\section{Observations and Data Reduction}

All of our mid-IR observations were obtained with the Infrared Array 
Camera (IRAC; \citealt{fazio04}) on the Spitzer Space Telescope 
\citep{werner04}. 
A log of the observations can be found in Table \ref{t:obslog}.  
The images were reduced and flux-calibrated with the S12 version of 
the Spitzer IRAC pipeline, and downloaded from the archive as 
Basic Calibrated Data (BCD). 
We obtained five BCD frames per object per wavelength channel, dithered 
with the medium-scale gaussian pattern. 
The IRAC channels have central wavelengths of 3.6, 4.5, 5.8, and 8.0 
$\mu$m for channels 1--4, respectively. 
The BCD images were corrected for array-location-dependency using 
the correction frames provided by the Spitzer Science Center. 
The frames were then combined with dual-outlier rejection to produce 
a mosaicked image for each channel using the Spitzer Science Center 
mosaicking software, MOPEX\footnote{i.e., Mosaicking and Point-source 
Extraction; see \citet{makovoz05} and the IRAC Data Handbook 
for a comprehensive discussion of the MOPEX package and the 
dual-outlier rejection algorithm.}.

\begin{figure}
\plotone{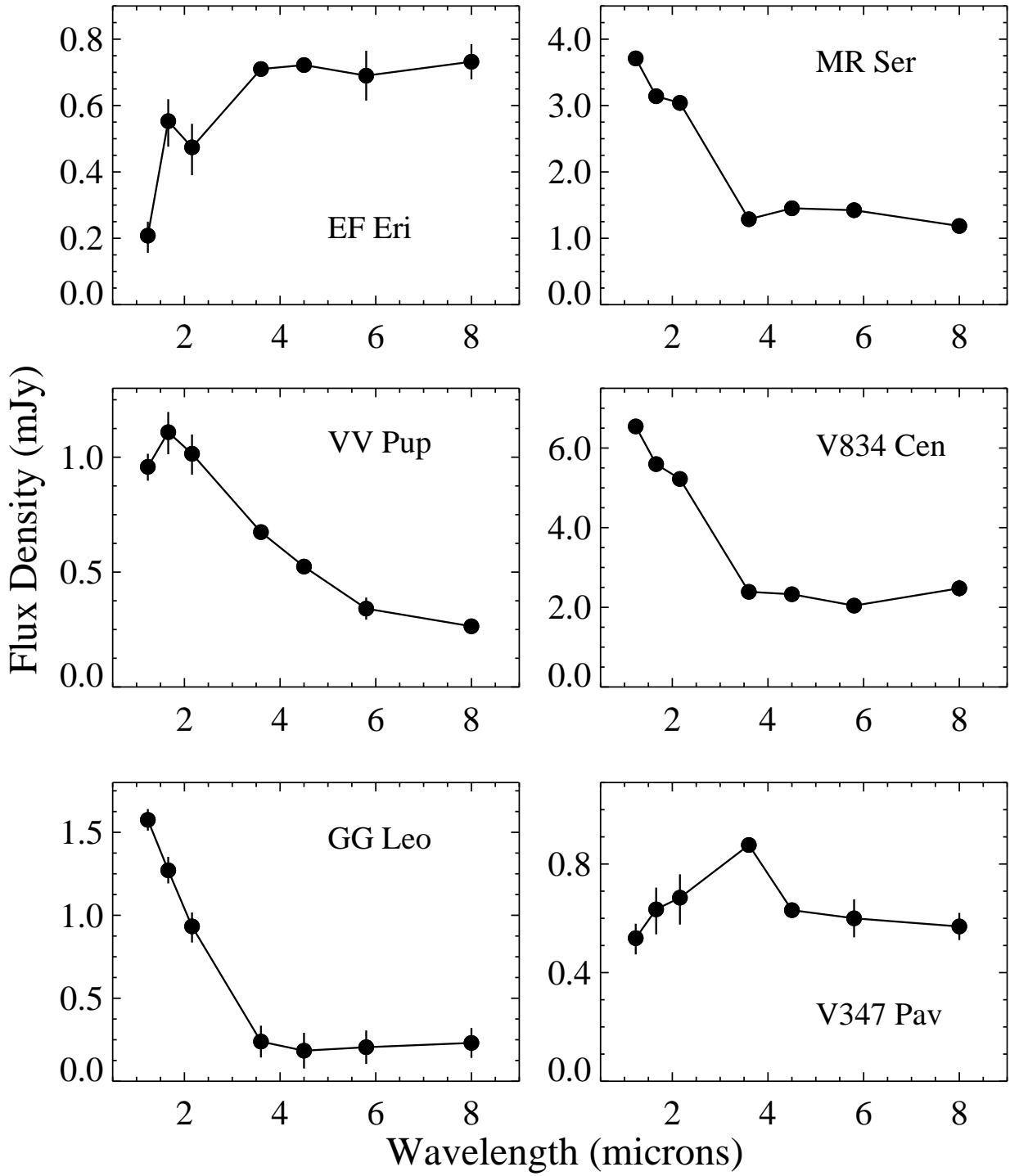}
\caption{Observed spectral energy distributions for our six polars, 
from 2MASS to IRAC.
\label{f:seds}}
\end{figure}

We then performed IRAF\footnote{The Image Reduction and Analysis 
Facility is maintained and distributed by the National Optical 
Astronomy Observatory.} aperture photometry on the mosaicked fields 
using an aperture size of 2.5 pixels.  
These results were subsequently corrected to an aperture size of 
10 pixels using aperture corrections derived from 46 isolated 
sources in our fields. 
The values of the aperture correction factors are 1.190, 1.199, 
1.211, and 1.295 for channels 1--4, respectively. 
From experience we found that the point response function (PRF) 
for the IRAC arrays is not well-modeled and we obtained more 
consistent results using aperture photometry than via PRF fitting. 
Sky subtraction was tested in three different ways:\ 
(1) using a sky annulus around the target; 
(2) using a sky region a short distance from the target (only for 
the more crowded fields of EF Eri and VV Pup); and, 
(3) subtracting a median sky from the whole mosaic before performing 
the photometry. 
The three methods were found to be consistent with each other. 
We extended the wavelength coverage for our data set into the near-IR 
using archival data from the 2MASS All Sky Point Source Catalogue 
\citep{skrutskie06}. 
The 2MASS $J$, $H$, and $K_{\rm s}$ magnitudes were converted to 
flux density in mJy using the absolute calibration zero 
points given in \citet{cohen03}.
All of the target photometry can be found in Table \ref{t:fluxes} 
and the resultant spectral energy distributions (SEDs) are
shown in Figure \ref{f:seds}.

\subsection{Error Analysis}

The only published attempts at rigorous error analysis for IRAC 
photometry are by \citet{lacy05}, who derived absolute errors of 
7\%, 12\%, 11\%, and 12\% for channels 1--4, respectively, and 
by \citet{reach05}, who carried out an absolute calibration of 
the IRAC instrument.  
The analysis by \citet{lacy05} was calibrated relative to 
corresponding 2MASS flux densities in an effort to obtain an 
absolute uncertainty for the IRAC colors. 
Those results therefore incorporate the uncertainties in both the 
2MASS photometry and the IRAC colors for their objects.  
As the absolute calibration for IRAC is likely to change as more 
data are obtained, we performed our own error analysis to determine 
the relative errors for our data, which can then be adjusted in the
future to incorporate the absolute errors as they become better known.

\begin{figure}
\plotone{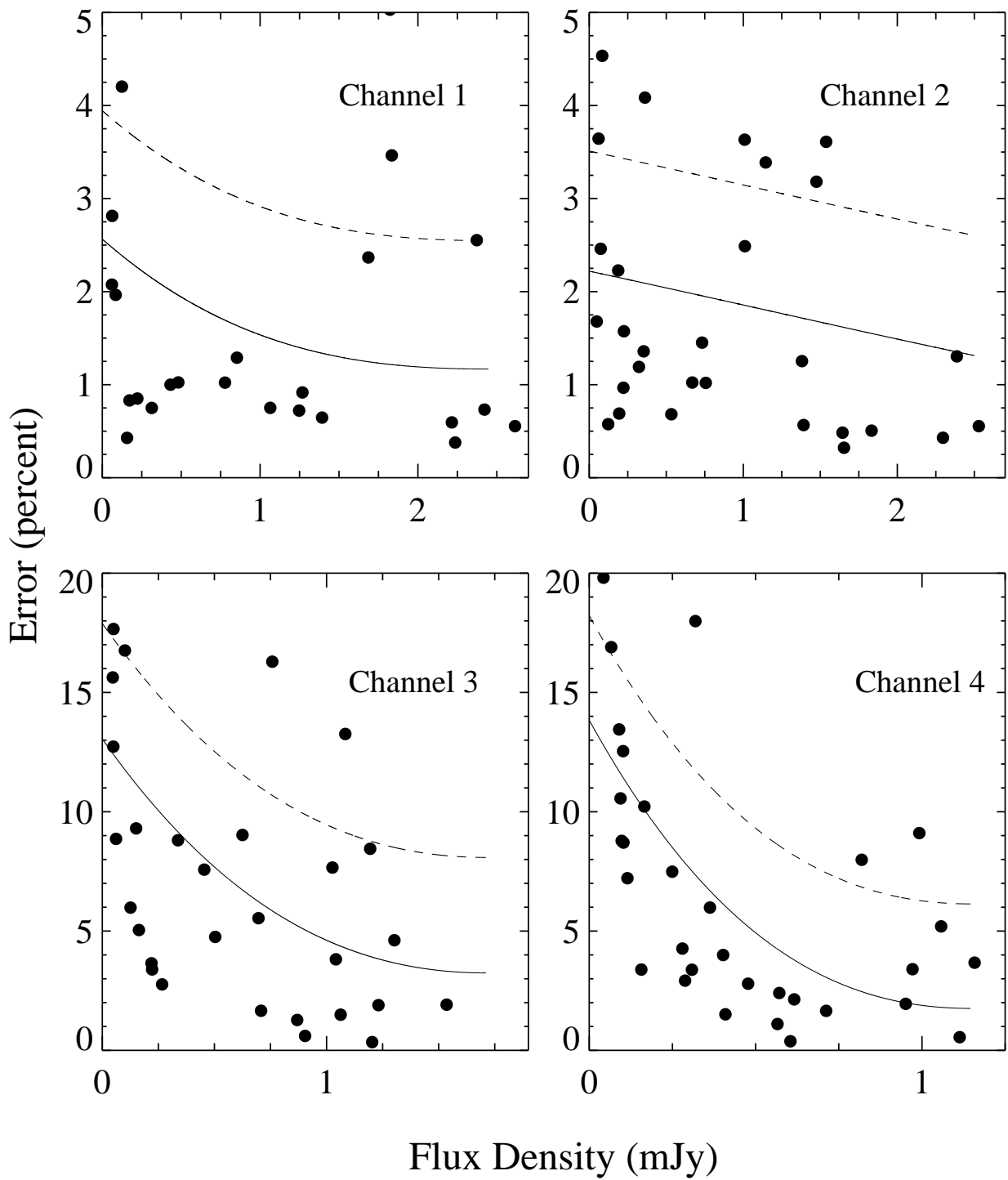}
\caption{Standard deviation of the flux densities in the five BCD 
frames vs. flux density measured in the mosaic for a number of 
isolated stars in our IRAC fields. Solid line is a best-fit 
polynomial to the data.  Dashed line is best fit + rms of residuals 
(i.e., the 1$\sigma$ uncertainty on the fit).\label{f:errors}}
\end{figure}

We performed aperture photometry on a number of isolated stars in 
each field (including fields for targets other than our polars), 
covering a wide range of flux densities. 
This was carried out for both the mosaicked images and the five 
individual BCD frames that were used to create each mosaic. 
The standard deviation of the five individual BCD flux density 
measurements of each test star was then plotted against the final 
flux density of the star measured from the mosaic 
(see Figure \ref{f:errors}), in the expectation that the standard 
deviation would decrease with increasing brightness. 
We find that there is indeed a correlation between the standard 
deviation and the final flux density, but there is considerable 
scatter around the best polynomial fit to the data (solid lines in 
Figure \ref{f:errors}). 
To allow for this we subtracted the polynomial fit from the data, 
calculated the standard deviation of the residuals, and added this 
to the best fit at each flux density to derive the total error 
estimate as a function of flux density for each IRAC channel 
(dashed lines in Figure \ref{f:errors}). 
We adopt these uncertainties for all future IRAC flux density measurements.

\section{Infrared Spectral Energy Distribution Models}

In Paper I, we briefly discussed the likely contributions to the
IR SEDs of polars from a number of expected system components.  In this
paper, we present detailed physical models to compare with our
observed SEDs.  The available model components are described below, and we
discuss the modeling results in \S\ref{s:modres}.

\subsection{Initial Values of Model Parameters}
\label{s:init_model_values}

\begin{figure}
\plotone{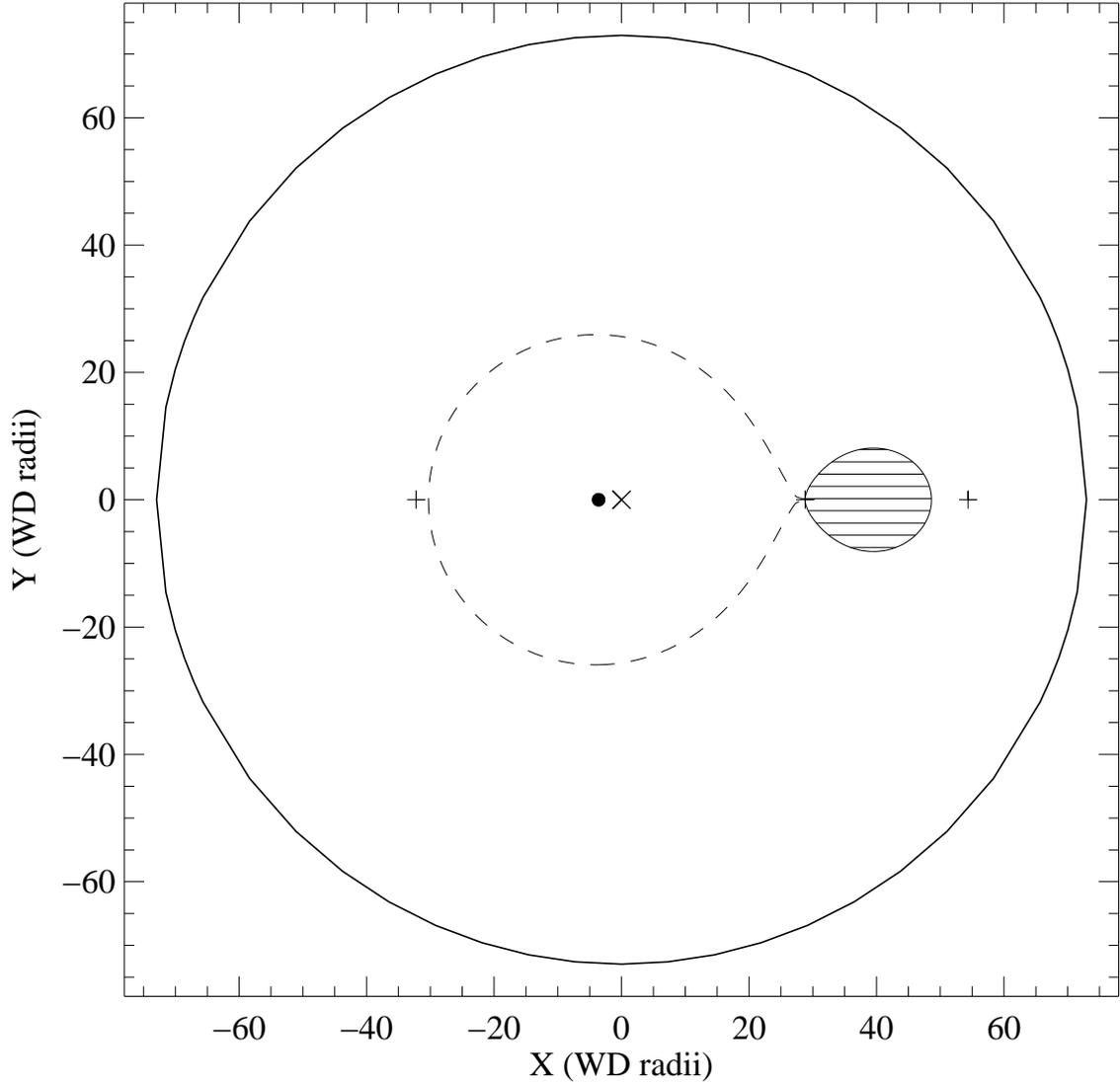}
\caption{Sample model system geometry of EF Eri (face-on, to scale) 
for the model shown in Figure \ref{f:eferi_unscaled}a and described 
in the text.
The center of mass (marked with an X) is at the origin and the outer 
Lagrangian points are shown as plus signs.  The WD is the small 
filled circle and the secondary star is the large hashed area.  
The secondary star fills its Roche lobe; the Roche lobe of the WD 
is shown as a dashed line.  The solid circle surrounding the CV is 
the inner edge of the circumbinary disk.
\label{f:eferi_geometry}}
\end{figure}

Our basic model includes several components (see below), 
each specified by a number of 
parameters (e.g., size, temperature, mass, etc.).  
We have held constant 
the values of as many as possible of these parameters using information
obtained from various literature sources for each target (see 
Table \ref{t:parameters}).  
Known relationships between physical parameters were used to infer the 
values of additional parameters from an observationally determined one; 
for example, given a WD mass based on observational data, we used the 
mass-radius relation for a magnetic, carbon-core WD with $\gamma=0.1$ 
from \citet{suh00} to calculate the corresponding WD radius.  
Radii and temperatures corresponding to a given secondary star 
mass and spectral type were obtained from \citet{cox00} and \citet{pont05}.  
In the absence of measured physical parameters in the literature, 
sub-stellar (i.e., L type) secondary stars were assumed to have 
$M_{2}=0.08M_{\odot}$, $R_{2}=0.1R_{\odot}$, and 
$T_{2}=$1500--2000 K (with higher temperature 
corresponding to earlier L spectral type; \citealt{kirkpatrick99}).  
In addition, parameters relating to the geometry of the CV (e.g., 
binary separation, Roche surfaces, etc.) were calculated via the 
well-known Roche potential relations (e.g., 
\citealt{kopal59, kopal69, kopal70}) that depend primarily on 
component masses and the orbital period, and can be solved numerically 
by utilizing an iterative approach (see Figure \ref{f:eferi_geometry}). 

The masses and radii of the secondary stars in our models are, to 
first approximation, consistent with filling their Roche lobes; 
however, whether or not the Roche lobe is filled is extremely sensitive 
to small changes in the mass or radius of both the WD and secondary star.
We expect that the masses and radii of the component stars, which we 
have obtained from the literature or calculated as described above, 
are not exactly known.  
Consequently, we have generally not been overly concerned with ensuring
that the model secondary stars precisely fill their calculated Roche lobes
for a given CV system geometry.
In a couple of cases, the secondary star spectral type required in the 
model that 
best reproduces the observed SED is not consistent with the mass 
estimates from the literature. 
In these cases, we suggest that either the mass estimates are inaccurate 
or we are encountering the well-known issue of the wide range of 
observed masses for each spectral type of CV secondary star 
(e.g., \citealt{smith98}).

\subsection{The White Dwarf}

The contribution of the WD primary star is simulated using a simple 
blackbody curve corresponding to the temperature of the WD,
scaled to the WD size and distance to the CV.  
Although, in detail, the SEDs of WDs are not identical to those of 
blackbodies, the falling Rayleigh-Jeans tail of the 
blackbody curve in the IR is a reasonable approximation to the long 
wavelength SED of a WD.  
In any case, the largest WD contribution is typically only a few per cent 
of the total observed flux density at 
the shortest IR wavelength (i.e., 2MASS $J$ band), and falls off 
rapidly to longer wavelengths.

\subsection{The Secondary Star}

The contribution of the secondary star is represented by choosing an 
empirical infrared SED template composed of $JHK_{\rm s}$ (2MASS) and 
3.6, 4.5, 5.8, and 8.0 $\mu$m (IRAC) observations of a number of 
M, L, and T dwarfs \citep{patten06}.  
The Patten et al.\ sample includes stars of the following spectral 
types:\ M3.5, M5, M5.5, M6.5, M7, M8, M9, L0, L1, L2, L3, L4.5, L6, 
L8, T0.5, T4.5, T5, T7, and T8. We caution that the sample includes only field stars, rather than stars in binaries that are subject to irradiation and mass-transfer effects, and so may not exactly reliably reproduce the SED of a CV secondary star. In comparisons made, however, it has generally been shown that CV secondary stars do not deviate significantly from their isolated counterparts \citep[e.g.][]{smith98}, and there is a long history of using field dwarfs to approximate the secondary stars in CVs. We also point out that no models of CV secondary stars exist at IRAC wavelengths, and therefore we use these templates as a first approximation.
We compiled the distances to each of the template stars from
various literature sources in order 
to transform all of their observed flux densities to a standard 
distance of 10 pc.  
Almost all of the template stars have distances measured via 
trigonometric parallax.  
The most distant template star is located at $\approx27$ pc, but 
60\% of them are closer than 10 pc.

We then performed a linear interpolation between adjacent observed 
spectral types in the M, L, and T groups, separately at each of the 
seven wavelengths, in order to ``fill-in'' the SEDs of the spectral 
sequence in steps of 0.5 in spectral type.  
The observed data for spectral types of M5.5, L2, and T5 from the 
Patten et al.\ sample were not used in the interpolation process 
because they showed anomolous flux densities at all wavelengths 
compared to the relatively smooth transitions between adjacent 
spectral types seen for the remaining stars.  
Possible explanations for this include errors in the published 
distances or spectral type classifications for these stars, or 
they may posssibly just be anomolous examples of their spectral type.  
We used interpolated SEDs to replace the observed SEDs for the 
three anomolous stars.

In all cases, \citet{patten06} estimate the uncertainties in the 
template flux densities as less than 2\%.  
Transformation to the standard distance will have increased this 
uncertainty by a few per cent, up to $\approx$10--15\% in the worst case.
All of our final model flux densities will be transformed to $d=10$ pc 
using the distances to each of the polars (either as a fixed, known 
value or as a free model parameter -- see \S\ref{s:init_model_values}) 
in order to utilize these empirical low mass stellar template SEDs 
as a model component.

\subsection{Cyclotron Emission}
\label{s:cyclotron}

As described in Paper I, we expect from a theoretical basis that 
cyclotron emission should be relatively unimportant as a contributor 
to the mid-IR SEDs of our polars, especially at the longest wavelengths.  
There is also empirical support for this claim.  
For example, \citet{dhillon98} found ``no evidence of cyclotron humps'' 
in the $K$-band IR spectrum of MR Ser.  The ``worst case'' scenario 
occurs for EF Eri, which has the lowest magnetic field strength of 
our sample of polars, with the cyclotron fundamental occuring at 
$\lambda\approx8$ $\mu$m.  Yet, \citet{ferrario96} found only very 
weak (amplitude $< 1$ mJy) cyclotron features at the red end of 
their 1--2.4 $\mu$m spectrum of EF Eri, a result that is also supported 
by the $K$-band spectra of \citet{harrison04}, who found only a very 
weak cyclotron hump longward of 2 $\mu$m, with stronger features only 
at shorter wavelengths.

In order to include a model component representing cyclotron emission, 
we first calculated a representative profile as a function of wavelength.
The shape of the cyclotron emission profile for a given field strength 
(which is a sum over harmonics) initially rises as $\nu^2$ before 
falling off rapidly with increasing frequency after the transition from 
optically thick to optically thin at some harmonic $1 < m < 10$ 
\citep{lamb79, cropper90, FKR02}. 
All of our polars except EF Eri have the cyclotron fundamental shortward 
of $\lambda\approx$ 4--6 $\mu$m, 
and we included the first 20 harmonics (which extends well into the optical 
regime in all cases).
The corresponding profile of the cyclotron emission is strongly affected 
by a change in the transition harmonic from a value of $m=1$, 
but the difference in the observed profiles for values of $m\geq2$
is relatively small. 
In general, the wavelength of the peak of the cyclotron emission component 
shifts shortward in proportion to $1/m$ as $m$ increases. 
So, at small (large) values of $m$, there is more cyclotron emission at 
longer (shorter) wavelengths in the model SED.
When $m\gtrsim3$, we expect to see only the initial rising (toward shorter 
wavelength) power law part of the cyclotron profile in the mid-IR SEDs 
of our polars.
We will investigate two cyclotron emission scenarios in our models, 
one in which the transition harmonic is at $m=1$ or 2 (i.e., approximately 
completely optically thin emission) and another in which $m\geq3$ 
(i.e., partially optically thick emission).

Electrons approaching the WD experience a successively stronger magnetic 
field as their distance decreases.  
Thus, it is more appropriate to consider the cyclotron emission as a 
sum over emission components for a range of field strengths.  
In order to approximate this effect, we recalculated the cyclotron 
profile in 1 MG steps from 1 MG to the specified surface field strength.  
Each profile was scaled by the distance-dependent factor described in 
\S3.1 of Paper I (to account for the change in electron density and 
temperature with distance from the WD).  
All of the profiles were then added together to obtain the final 
cyclotron emission profile for a given field. 
For the same surface field strength ($B$), the ``sum over field strengths'' 
prescription generally results in
more cyclotron flux at longer wavelengths than in the single-field case. 
It also produces a cyclotron ``pseudo-continuum,'' which tends to 
smooth over individual cyclotron humps (we already do not expect to see 
a strong signature of cyclotron humps in our model SEDs due to the poor 
spectral resolution of our data, which is matched in the model 
calculations).

The final parameter for the cyclotron component is a scale factor, 
which is multiplied with the model profile to match the observed 
flux densities.  
This model component is the only non-physical component, in the 
sense that the absolute flux level is not predicted solely from 
other physical parameters (like temperature, distance, etc.).  
However, we are able to strongly constrain the scale factor for 
the cyclotron component, because when it is combined with the 
WD and secondary star components, the total cannot exceed any of the 
observed flux densities.

\subsection{The Circumbinary Dust Disk}
\label{s:cbd}

The presence of a combination of optically thick and optically thin 
dust was first proposed by \citet{berriman85} as an explanation for 
the IR luminosity in CVs.  Initial evidence for the likely presence 
of dust in several polars observed with Spitzer was described by us 
in Paper I.  We now approach this analysis in more detail by including 
a realistic estimate of dust emission as a component in our SED model.

There are several possible origins for dust in CVs, including 
(1) remnant dust produced during the common envelope phase and ejected 
from the binary with some angular momentum, 
(2) dust lost from the accretion flow and ejected from the binary, 
(3) dust produced in the mass outflows during repeated nova or dwarf 
nova outbursts that does not escape from the gravity of the binary 
(e.g., \citealt{ciardi06}), 
and/or (4) dust produced over long timescales in the mass outflow of 
a wind from the accretion disk and/or WD 
(see \citealt{taam01}, \citealt{spruit01}, 
\citealt{taam03}, \citealt{dubus04}, and \citealt{willems05} for further 
discussion of the possible origin of circumbinary dust in CVs).  

Dust that is located outside the Roche lobes of the stellar 
components in a CV should settle into a relatively cool, 
geometrically thin disk \citep{spruit01, taam03, belle04, dubus04}. 
There is a growing body of observational evidence for circumbinary 
dust disks in both post-AGB binary systems 
\citep{vanwinckel04, deruyter05, deruyter06} 
and pre-main-sequence binary stars \citep{hartmann05}. 
The detection of crystalline silicates in these circumbinary dust disks 
has been interpreted as an indicator of the long lifetimes of the 
disks \citep{hartmann05, deroo06}. As there is no previous observational evidence of circumbinary dust disks around CVs, it is impossible to say whether they have a similar longevity. However, the assumed temperature and density of the circumbinary disks are approximately an order of magnitude too low for significant sputtering of the grains \citep{burke74}, and  if we assume that they are similar to those observed around other binary systems,
then the dust disks will require little or no replenishment.  The 
total minimum mass of dust that must be ejected from the system in 
order to form a circumbinary disk is therefore relatively small 
(approximately the mass of a large asteroid).  
This assumes that the dust is formed in the inner system prior to 
ejection, most likely in the outer atmosphere or wind 
of the low-mass secondary 
star \citep{lunine89, allard01, leggett02, tsuji02} -- the probability 
of the dust forming in-situ in the disk itself is extremely low, 
as the dust formation mechanism requires high gas densities and, 
therefore, would require a high mass transfer rate from the inner 
binary system into the disk \citep{berriman85}.

Recently, \citet{muno06} reported the discovery of a Spitzer/IRAC 
mid-IR excess around several low mass X-ray binaries (LMXBs), which 
they interpret as evidence for circumbinary dust disks with almost 
identical characteristics to the ones we model in this paper. 
Searches for dust disks around CVs have so far failed to uncover 
any direct evidence for their existence (e.g., \citealt{belle04}).
In most CVs the circumbinary dust disk flux 
in the IR (if present) is likely to be overwhelmed by radiation from the cool 
outer regions of the accretion disk (or even the bright, early spectral 
type secondary star in long orbital period systems). 
In that regard, polars may be the only type of CV 
in which the flux contribution from cool dust can be detected. 
Despite the lack of direct observational evidence, the existence 
of circumbinary disks in CVs has been suggested as the explanation 
for both the high mass transfer rate in some systems, as well as 
the observed lack of pile-up of systems with orbital periods near 
the period minimum \citep{spruit01, taam01, taam03, willems05}.
However, as described in more detail in \S\ref{s:discussion}, the 
estimated total masses of our model circumbinary disks are likely 
too small to influence CV evolution as described in the papers 
cited above.

For inclusion as a component in our SED model, we will initially 
make the assumption that the circumbinary dust disk is optically 
thick, as this presents a computationally simpler case, and even 
(as described in \citealt{dubus04}) might be the most appropriate 
case for dust disks in CVs.  For completeness, however, 
we will also consider the case of an optically thin dust disk.

\subsubsection{The Optically Thick Case}

Our optically thick, geometrically thin dust disk model for the 
polars is based on the description of the expected properties of 
circumbinary disks in CVs given in \citet{dubus04}. We do not consider the case for a warped or flared dust disk as our small number of data points cannot support the extra parameters required to model anything more complex than the geometrically thin case. 
We utilize the analytic solution for an optically thick dust 
disk radial temperature profile and corresponding integrated flux 
density derived in \citet{jura03}, \citet{becklin05}, and 
references therein.  
In particular, equation 1 from \citet{jura03}
provides the temperature profile of the disk:\ 

\begin{equation}
T_{\rm disk} \approx \left(\frac{2}{3\pi}\right)^{1/4} 
\left(\frac{R_{\ast}}{r}\right)^{3/4} T_{\ast},
\label{e:disktemp}
\end{equation}

\noindent where $T_{\ast}$ and $R_{\ast}$ are the temperature and radius 
of the central irradiating star, respectively, and $T_{\rm disk}$ 
is the resultant temperature of the disk at radial distance $r$.
Equation 2 from \cite{jura03} provides the integrated flux 
density, $F_{\rm disk}$:\

\begin{equation}
F_{\rm disk} = \frac{2\pi\cos i}{d^{2}} 
\int_{r_{\rm in}}^{r_{\rm out}}B_{\nu}(T_{\rm disk})r dr,
\label{e-diskflux}
\end{equation}

\noindent where $i$ and $d$ are the system inclination and distance, 
respectively, 
$r_{\rm in}$ and $r_{\rm out}$ are the inner and outer radii of the 
disk, and $B_{\nu}$ is the Planck function.
In this formulation, the dust disk is assumed to be optically thick 
but geometrically thin (i.e., with negligible height), so the 
integrated flux density originates only from the visible surface of 
the disk (with no contribution from the negligibly tall disk rim).  

In principle, the optically thick dust model is not sensitive 
to the exact properties of the dust grains (size, composition, etc.), 
as long as the grains are assumed to uniformly (re)radiate as blackbodies.  
The exponent on the radius-dependence in the equation for the disk 
radial temperature profile, however, does have some implication for 
the assumed properties of the dust grains. 
A large exponent leads to lower temperatures near the inner edge of 
the disk, which corresponds to large dust grains that cool efficiently.
Conversely, a small exponent leads to higher temperatures near the 
outer edge of the disk, which corresponds to small dust grains that 
do not cool efficiently.
In practice, we have found that an exponent of 3/4, as used in 
equation \ref{e:disktemp}, is the only value that produces viable 
results.  Larger exponents produce disk SEDs that rise too shallowly 
at longer wavelengths and are, overall, too faint to match the 
observed mid-IR flux densities without arbitrarily increasing the 
temperature of the disk's inner edge far beyond what could be 
produced by the irradiating stars.  Smaller exponents produce disk 
SEDs that rise too steeply and are, overall, too bright to reproduce 
the observed SEDs without arbitrarily increasing the distances to 
the CVs to many hundreds or thousands of pc. Departure from an exponent 
of 3/4 implies non-blackbody grains, which are further discussed in 
Section 3.5.2.

The disks considered by \citet{jura03} and \citet{becklin05} are 
around isolated WDs (G29-38 and GD362, respectively).  
There are a number of differences that must be considered when 
applying this model to a circumbinary disk in a CV system.  
First, the disk will be centered around the center-of-mass of the 
CV instead of the WD.  In the case of a low mass (potentially 
sub-stellar) secondary star, the difference between the center 
of the WD and the center-of-mass of the CV is relatively small 
($\lesssim$ 10\% of the binary separation).  
Second, the inner edge of the disk will be much further from the WD.
In the WD disk models, the inner edge of the disk is typically 
located at a few to $\sim10$ WD radii.  However, in the case of a 
circumbinary disk, the inner edge is, by definition, located outside 
the Roche lobes of the stellar components.  We assume an inner radius 
for the circumbinary disk equal to the tidal truncation radius at 
1.7 times the separation of the stellar components in the CV 
\citep{dubus04}.  For our short-orbital-period polars, this 
typically corresponds to $r_{\rm in} \gtrsim 100R_{\rm WD}$.  
The outer edge of the disk is calculated to correspond to a 
temperature of 20 K using equation \ref{e:disktemp}. This typically 
corresponds to $\sim$0.5--2$\times10^{4}$ WD radii for our polars, but in reality, the radius of the outer disk has a negligible effect on the model flux densities at the IRAC wavelengths, and serves only as a conservative, consistent cut-off value. Third, we must consider irradiation heating of the disk from not 
only the WD but also the secondary star\footnote{Fortunately, since 
polars lack an accretion disk, we do not have to consider this 
additional potential irradiating source.  In principle, the 
accretion spot(s) at the WD magnetic pole(s) could be considered 
as additional discrete sources of irradiation; however, unless the 
magnetic axis is aligned very close to the WD equator, the 
geometrically thin circumbinary disk will not ``see'' the geometrically small 
accretion spot(s).}.  For example, for a typical short-period CV 
composed of a WD with temperature $T_{\rm WD} = 10,000$ K and 
radius $R_{\rm WD}=0.01R_{\odot}$, and a low mass secondary star 
with $T_{2} = 3000$ K (i.e., $0.3T_{\rm WD}$) and radius 
$R_{2} = 0.1R_{\odot}$ (i.e., $10R_{\rm WD}$), the ratio of total 
luminosities (given by $L \propto R^{2}T^{4}$) is 
$L_{2}/L_{\rm WD} = 0.8$.  Thus, the total luminosity of the 
secondary star is comparable to that of the WD.  

Unfortunately, the additional contribution from the secondary star 
makes our consideration of the heating of the circumbinary disk 
more complex, because the secondary star is much further from the 
CV's center-of-mass (i.e., the center of the circumbinary disk) 
than the WD, so is, at any instant, closer to the inner edge of 
the circumbinary disk.  However, under the assumptions that the 
disk inner edge radius is large compared to the binary separation 
(see above) and that the orbital period is short (so, on a 
time-averaged basis, all parts of the disk ``see'' the secondary 
star equally), we will neglect the offset of the secondary star and 
assume that all disk irradiating sources are located at the 
center-of-mass.

Using the typical CV parameters listed above with $P_{\rm orb}=90$ min,
and stellar component masses of 
$M_{\rm WD}=0.6M_{\odot}$ and $M_{2}=0.1M_{\odot}$, the 
circumbinary disk will have an inner edge radius of 
$R_{\rm in}=100R_{\rm WD}$.  Assuming only heating from the WD, 
equation \ref{e:disktemp} predicts an inner edge temperature of 
$T_{\rm in}\approx200$ K.  If only the secondary star is 
considered as the heating source, then we get $T_{\rm in}\approx350$ K.  
We have considered two methods for estimating the temperature 
of the circumbinary disk's inner edge that is likely to result 
from heating by {\em both} stellar components in the CV.
The first method is to simply add together the inner edge 
temperatures predicted by heating from each of the stellar 
components individually.  This would give $T_{\rm in}\approx550$ K 
for the typical CV parameters described above.  
The second method is to calculate a single ``equivalent'' star with 
radius $R_{\rm equiv}$ that is equal to the quadrature sum of 
$R_{\rm WD}$ and $R_{\rm 2}$, and temperature $T_{\rm equiv}$ such 
that $L_{\rm equiv}=R_{\rm equiv}^{2}T_{\rm equiv}^{4}$ is equal 
to the sum of the individual luminosities of the WD and secondary star.  
For the system parameters listed above, this gives 
$T_{\rm in}\approx450$ K.
Given the uncertainty in estimating the likely heating effect of 
both stars, and the fact that both of our suggested methods yield 
similar temperatures, we have used the circumbinary disk inner 
temperature as a free parameter in our model SED calculations.  
However, we have constrained $T_{\rm in}$ to be within 
$\approx \pm200$ K of the value predicted for heating by a single ``equivalent'' star 
(i.e., the second method described above).

\subsubsection{The Optically Thin Case}

The basic structure of an optically thin circumbinary dust 
disk is identical to that in the optically thick case -- 
it is geometrically thin and heated by irradiation from both stellar 
components, both of which are assumed to be located at the system 
center-of-mass. Our model is based on two main assumptions -- 
first, that 
each dust particle (re)radiates as a blackbody and, second, that all of
the particles are ``seen'' by the observer (i.e., there is no eclipsing 
of the disk by the stellar components and every one of the grains 
contributes to the total flux density).  The former condition is not an 
unreasonable assumption since 
the inner disk radius is set at the tidal truncation radius, well away 
from the stellar components. 
We use the same disk temperature profile 
from \citet{jura03} and \citet{becklin05} as we use in the optically 
thick case, with an outer disk radius corresponding to an outer disk 
temperature of $T_{\rm out} = 20$ K, also as used in the optically thick 
case. We assume that the disk has a constant volume density 
throughout, so the volume of material at any temperature (radius) 
in the disk increases with falling temperature (increasing radius). 
We used a fixed grain radius of 10 $\mu$m, which was suggested by 
\citet{jura03} as the minimum grain size for dust debris disks around 
main sequence stars, and a fixed dust grain density 
of 3\,g\,cm$^{-3}$.  In fact, the exact values of the dust
grain parameters that we use have little effect on 
the resultant model SED of the dust disk; for example, if we 
change the grain size (thereby reducing the available radiating
surface area in the disk), then the required luminosity 
to match the observed SED can still be achieved 
by changing one of the other disk 
parameters, such as the volume density of the dust. 

In all cases, we kept the disk thickness and volume density as low as 
possible to ensure that the disk remained optically thin; however, 
this was very difficult to quantify. The optical depth of the 
dust is given by: 

\begin{equation}
\tau = \kappa_{\rm s}\rho S,
\end{equation}

\noindent where $\kappa_{\rm s}$ is the mass absorption coefficient, 
$\rho$ is the 
volume density of the dust, and $S$ is the column length along the 
line of sight. The mass absorption coefficient is a 
function of dust composition, grain size and shape, dust 
temperature, and the frequency of the incident radiation. As a 
result, it is usually quoted in a frequency-dependent form, 
$\kappa_{\nu}$. It is highly temperature-sensitive, but extremely 
poorly studied for temperatures above $\sim$300K, as all of the 
literature on this subject deals with the cool interstellar medium 
and there are no published values of $\kappa_{\nu}$ at the 
temperatures found in our disks. Even the theoretically 
calculated values for the relatively well-studied low temperatures 
vary by a factor of 30, from 0.002 to 0.10 cm$^2$ g$^{-1}$ 
\citep{draine89}. As a result, we have no certain way of testing 
the optical depth of our disks, but we adopted a simple, generic 
relation in an attempt to quantify our values:

\begin{equation}
\kappa_{\nu} = 0.1(\nu / 10^{12} \, {\rm Hz})^{\beta} \, {\rm cm}^2 \, 
{\rm g}^{-1}, 
\end{equation}

\noindent which is the general form given in \citet{beckwith90}. The value of $\beta$ depends on the growth shape of the grains - spheriodal grains will have a higher value of $\beta$ than fractal ones\citep[see][for an in-depth discussion of these effects]{beckwith90}. We follow the example of \citet{beckwith90} and accommodate the uncertainty in the grain shape by adopting the general relation given above for $\kappa_{\nu}$ with $\beta$=1.
In any case, the volume density, and therefore the optical depth, 
of the disk is dependent on the disk temperature profile, the 
grain size and mass density, and the inner and outer radii of the 
disk.  Any calculation of the true optical depth of the disk requires 
better constraints on all of these parameters.

When the presence of dust can explain the long wavelength end of a 
polar SED, we generally have been unable to differentiate between 
the optically thick and thin circumbinary dust disks -- because of 
the number of available parameters used to describe each type of 
disk, we can produce qualitatively similar model SEDs using either 
type of disk.  Thus, in the remainder of this paper we will 
generally only discuss the less complex optically thick case, with 
the exception that we will consider an estimate of the total dust 
mass from the optically thin case (see \S\ref{s:discussion}).

\section{Modeling Results}
\label{s:modres}

We emphasize that we have seven data points in 
each of our SEDs, 
only four of which (from IRAC) are in the mid-IR.
Consequently, our modeling results 
are intended to illustrate general trends in polars as a class 
rather than provide a specific, detailed representation of 
each of our target polars. 
In general, we started each model by including best estimates for
the WD and secondary 
star components, then attempted to fit the observed 8 $\mu$m flux 
density through the addition of either cyclotron or dust emission, 
without exceeding the observed flux density at any shorter wavelength.  
In reality, these systems could (and, likely, do) include 
contributions from {\em both} cyclotron and dust emission, but we 
have too few data points to constrain such a complex model in any 
meaningful way.  
However, as will be described in detail in the following sections, 
success at reproducing the observed
8 $\mu$m point with a particular model 
configuration while simultaneously not exceeding the observed 
flux densities at shorter wavelengths can allow us to rule out 
cyclotron emission (often) or dust (occasionally) as a dominant 
contributor at the long wavelength end of the observed SEDs.

Because of the relatively small number of observational constraints
compared to the degrees of freedom in the models, optimization of 
model parameters was accomplished via manual testing.  
In general, the values of the model parameters can be varied within 
a range of about $\pm$5--10\% and will still produce a model SED of 
comparable quality to the ``best'' cases discussed below.
For convenience in discussing our results, 
we divided the targets into three classes based 
on observed SED morphology.  
Class I contains EF Eri and V347 Pav, which have SEDs in which 
the IRAC portion is brighter than the 2MASS portion.  
Class II contains VV Pup, whose SED is most similar in shape to 
a falling Rayleigh-Jeans-like tail and may be dominated by the 
secondary star.
Class III contains V834 Cen, GG Leo, and MR Ser, which have SEDs 
in which the 2MASS portion is bright and falls steeply toward 
long wavelengths but the IRAC portion is relatively flat.
We will now discuss our modeling results for the polars in each 
of these classes.

\subsection{Class I -- Bright IRAC}
\label{s:class1}

The two polars in this class have 
oddly-shaped SEDs compared to the other four polars in our sample.
Specifically, the IRAC portion of the SED, while still approximately 
flat at the longest wavelengths, is brighter than the 2MASS portion 
(see Figure \ref{f:seds}).  
This possibly stems from another potential complication to the 
interpretation and modeling of our polar SEDs; namely, that polars  
are highly variable objects.  Their brightness changes dramatically 
on irregularly-spaced long-timescales (weeks to years) between high 
and low accretion states (by up to a few magnitudes), and over their 
orbital period (by as much as 2 mag). In order to best compare the 
non-simultaneous 2MASS and IRAC measurements for our targets, we 
attempted to determine both the accretion state and orbital phase 
of each of the polars at the times of both sets of measurements.  

We used flux measurements from the literature and long-term light 
curves from the American Association of Variable Star Observers 
(AAVSO) to check the accretion state of the polars.
Of our six targets, we were only able to definitively confirm that 
one (EF Eri) was in the same accretion state (low) during both the 
2MASS and Spitzer observations.  We suspect that at least three of 
the systems (V834 Cen, GG Leo, and MR Ser; i.e., the Class III 
objects) were in high accretion states during the 2MASS observations 
because of the steepness of the 2MASS portion of their SEDs, which 
may be indicative of the falling Rayleigh-Jeans-like tail of 
accretion-generated luminosity.  The 2MASS--IRAC SED transition in 
VV Pup (i.e., the Class II object) is very smooth, which suggests 
that the data were obtained during the same accretion state 
(regardless of which state).
Light curves from the AAVSO show that VV Pup was in a high state 
for the 2MASS observations. There is no ground-based coverage at 
the time of the Spitzer observations, but from the long-term light 
curve, it is obvious that the system spends most of its time in a 
high state, and the closest data points show no indication of the 
system dropping into a low state at that time.  

So far, there are no published orbital light curves of polars in the mid-IR, 
so we have no evidence that the flux levels at the IRAC wavelengths 
vary with the same phasing and/or amplitude as in the optical or 
2MASS wavebands, but for completeness we consider this possibility. 
We used published ephemerides to determine the orbital phasing 
of the 2MASS and IRAC data sets for our targets, and then used 
published optical or near-IR light curves to estimate whether the 
orbital brightness states of those data sets were comparable.
Both EF Eri and V347 Pav were observed by 2MASS and Spitzer at 
significantly different orbital phases that correspond to different 
orbital brightness states (in both cases, the 2MASS data were obtained 
during a faint state, and the IRAC data during a bright state).
The other four targets were observed by both 2MASS and Spitzer at 
similar orbital phases that correspond to the same orbital brightness 
state, and so the results at all wavelengths should be directly comparable.
In the absence of mid-IR orbital light curves, we have assumed that the 
ratio of bright-to-faint flux density for the IRAC data is the same as
observed in the optical/near-IR.  In any case, there are enough model
parameters to compensate for different bright/faint scale factors 
(and, of course, if the mid-IR is shown in the future to {\em not}
vary over the orbital period like the optical/near-IR, then this
contingency can be discarded!).
We note that even after attempting a scaling correction, the SEDs 
for EF Eri and V347 Pav are morphologically similar to each other 
but different from the other classes we have defined 
(see \S\ref{s:eferi} and \S\ref{s:v347pav} for details).  They 
show flat IRAC SEDs like Class III, but the 2MASS portions of 
their SEDs resemble Class II.

\subsubsection{EF Eridani}
\label{s:eferi}

We selected EF Eri for our IRAC campaign due, in large part, 
to its extremely red ($J-K_{\rm s} > 2.5$) 
2MASS colors \citep{hoard02}, which exceed even the reddest IR colors 
for mid-L dwarfs \citep{leggett02}. 
This CV was discovered as an X-ray source by \citet{gursky78} and 
subsequently identified by \citet{griffiths79} and \citet{williams79} 
as a probable AM Her type. 
The magnetic nature of the system was 
confirmed by \citet{tapia79} from measurements of the linear and 
circular polarization, which he used to derive the very short 
orbital period of P$_{\rm orb} \approx 81$ min, which was later 
refined by \citet{piirola87}.

EF Eri entered an extended low state 
in 1997, only emerging back into a high state in March 2006, and 
only remaining bright for a few weeks before dropping back into 
the low state. 
The low state optical spectrum obtained by \citet{wheatley98} 
displayed only weak line and cyclotron emission, indicating that 
the low state mass transfer rate is extremely small. 
Modeling of this spectrum by \citet{beuermann00} set a limit 
of M9V or later for the secondary star spectral type, and a distance
estimate of $d=100\pm20$ pc (consistent with the upper limit of $d=119$ pc
estimated from near-IR observations by \citealt{YS81}). 
\citet{howell01} detected the secondary star in $K$-band spectra 
with a spectral type of either M6V or L4--5, although it is 
impossible to tell whether the star has dropped below the hydrogen 
burning limit based solely on spectral type, as the temperatures 
may depend on irradiation and evolutionary history.  
\citet{harrison03} presented simultaneous multi-wavelength optical 
and infrared photometry in which they believed they detected a 
brown dwarf secondary star in the $H$ and $K$ bands, but follow-up 
phase-resolved spectroscopy \citep{harrison04} showed that the 
variability in both bands is dominated by cyclotron emission 
throughout the orbital cycle, most likely originating at both 
of the WD magnetic poles.  

Recently, \citet{howell06c} determined an extremely low mass for 
the secondary star in EF Eri ($M_2=0.055M_{\odot}$), suggesting 
that an L spectral type classification is, indeed, appropriate.
At a distance of $d \approx 100$ pc, the secondary star in EF Eri cannot 
be earlier than L5.0, or the combined flux densities of just 
the WD and secondary star will significantly exceed the observed $J$-band 
point.  
Our model for EF Eri uses the WD parameters from \citet{beuermann00}, 
the secondary star parameters from \citet{howell06c}, and 
assumes a spectral type of L5.0 for the secondary star (see Table 
\ref{t:parameters}).  We then combined this model with an additional 
component representing cyclotron or dust emission to explore the 
limiting cases for the ability of either mechanism to explain the 
observed long wavelength SED of EF Eri (without exceeding the 
short wavelength end).

\begin{figure}
\plotone{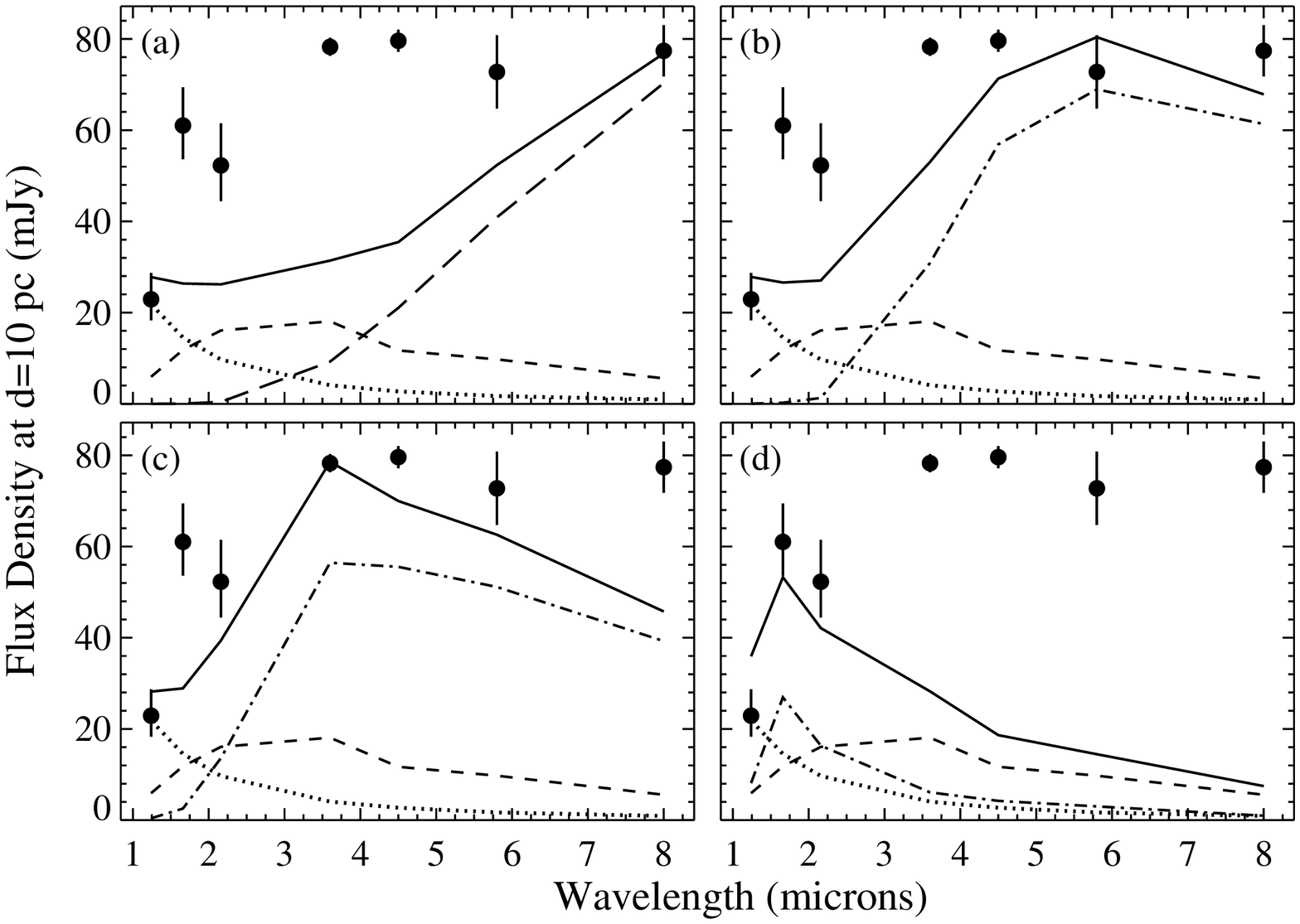}
\caption{Representative model SEDs for EF Eri showing the observed 
data (filled circles), total model SED (solid line) and model 
components: the WD (dotted line), the secondary star (short dashed 
line), a circumbinary dust disk (long dashed line), and cyclotron 
emission (dot-dashed line).  
The plot panels show (a) circumbinary disk model, (b) cyclotron 
model with $m=2$ and (c) $m=3$, and (d) single-field cyclotron 
model with $m=5$.
See text and Table \ref{t:eferi_parms} for details.
\label{f:eferi_unscaled}}
\end{figure}

Figure \ref{f:eferi_unscaled} shows representative model SEDs 
for EF Eri.  The model containing a circumbinary dust disk (Figure 
\ref{f:eferi_unscaled}a and Model 1 in Table \ref{t:eferi_parms}) 
can easily reproduce the observed 8 $\mu$m point, but significantly 
underestimates the observed flux densities at the shorter 
wavelengths (except at the shortest wavelength, $J$-band).  
When the cyclotron transition harmonic is set to $m=1$, 
the cyclotron emission is essentially indistinguishable from the 
circumbinary disk component, so, in the case of completely optically 
thin cyclotron emission, we cannot differentiate between cyclotron 
or dust producing the 8 $\mu$m emission.  However, neither mechanism 
(on its own) can produce the high levels of emission at shorter 
wavelengths without greatly exceeding the 8 $\mu$m point.  
We found this to be true in all of our polars, so we do not mention 
it again in the discussion of our modeling results, and defer 
discussion of completely optically thin cyclotron emission 
to \S\ref{s:discussion}.

Figure \ref{f:eferi_unscaled}b (see Model 2 in Table \ref{t:eferi_parms}) 
shows the model SED having a cyclotron component with transition 
harmonic at $m=2$, which best reproduces the observed long 
wavelength end of the SED, but still underestimates both the short 
wavelength end and the 8 $\mu$m point (even when the model 5.8 $\mu$m 
flux density is pushed to the upper uncertainty limit of the observed 
point at that wavelength, as shown in the figure).  Switching the 
cyclotron component to increasingly optically thick configurations 
(e.g., $m=3$; see Figure \ref{f:eferi_unscaled}c and Model 3 in 
Table \ref{t:eferi_parms}) recovers some of the short wavelength 
flux, but at the expense of substantially underestimating the IRAC 
portion of the SED.

Up to now, we were using the sum-over-field-strengths prescription 
for the cyclotron emission (see \S\ref{s:cyclotron}).  As shown in 
Figure \ref{f:eferi_unscaled}d (Model 4 in Table \ref{t:eferi_parms}), 
the single field strength approach can approximately reproduce the bright 
observed $H$ band point (at the expense of over-estimating the J-band point) because the cyclotron humps are not blended 
together into a ``pseudo-continuum'' as in the sum-over-field-strengths 
approach.  It is possible that if the 
intensity scaling of profiles produced further from the WD 
(i.e., at smaller magnetic field strengths) falls off more 
steeply than estimated in Paper I, then the final profile 
might be dominated by the emission produced very near the 
WD surface and, hence, retain the sharply peaked shape seen in 
Figure \ref{f:eferi_unscaled}d.
A better reproduction of the EF Eri SED might be achieved through 
a combination of models similar to Models 1 and 4 (i.e., the SEDs 
shown in Figures \ref{f:eferi_unscaled}a and -d).  However, even 
in this case, there would be some ``missing'' flux at 3.6 and 
4.5 $\mu$m, suggesting the need for either model components with 
different spectral profiles or an additional emitting system 
component.  Adding more model parameters/components is not 
warranted by the ability of these sparse data to constrain complex 
models.  We note, however, that some experimentation demonstrated 
to us that a simple blackbody component with $T\approx1000$ K 
produces an emission profile that peaks around 4.5 $\mu$m and 
could fill in the EF Eri model SED as required. 

\begin{figure}
\plotone{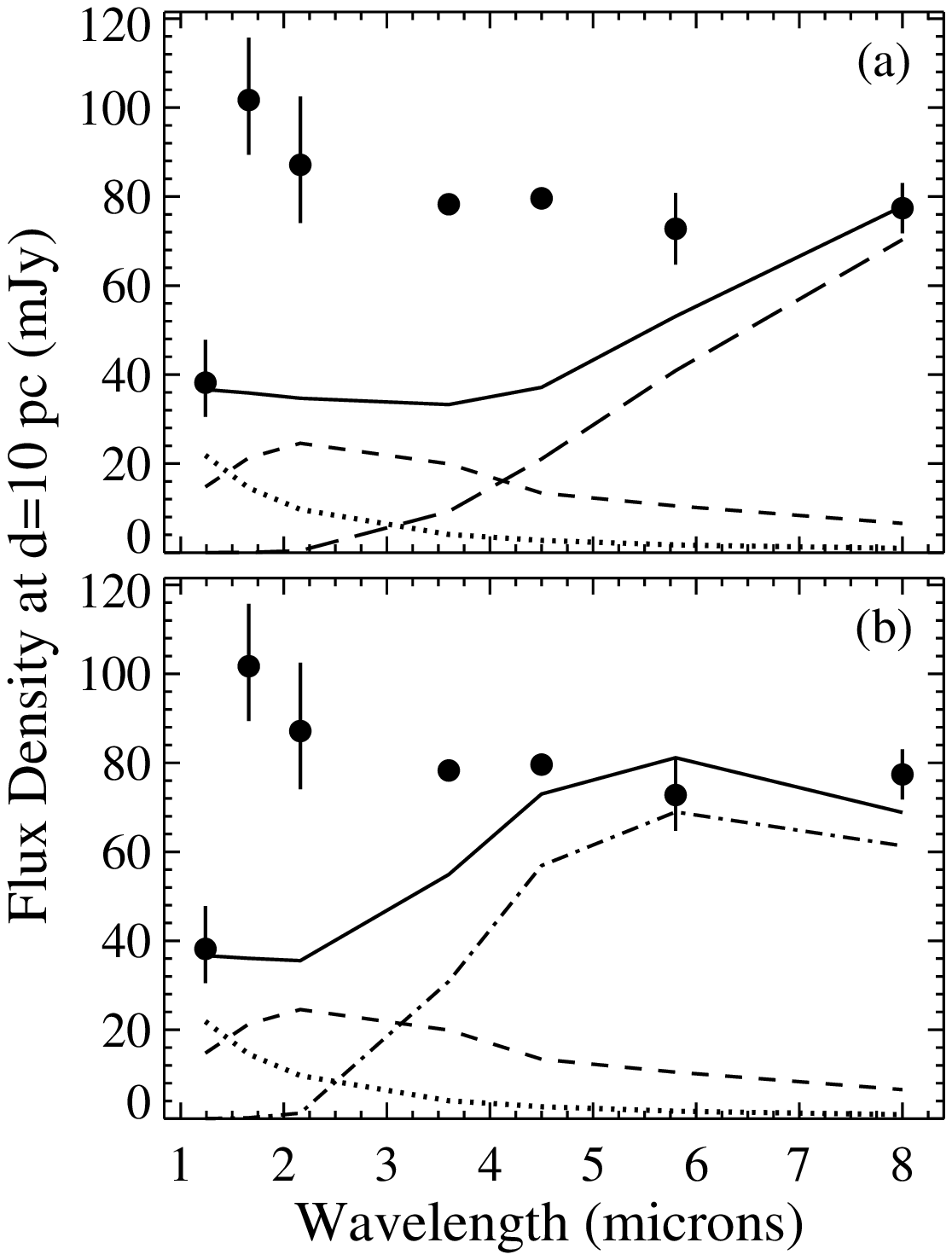}
\caption{As in Figure \ref{f:eferi_unscaled}, but with the 2MASS 
data for EF Eri scaled by a factor of 1.67 to account for potential 
orbital brightness variations.
The plot panels show (a) circumbinary disk model and (b) cyclotron 
model with $m=2$.
See text and Table \ref{t:eferi_parms} for details.
\label{f:eferi_scaled}}
\end{figure}

As noted in \S\ref{s:class1}, the 2MASS and IRAC data for EF Eri 
were obtained during different orbital brightness states (faint 
and bright, respectively).
The low-state $K$-band photometric light curves of \citet{harrison03} 
show that $F_{\rm bright}/F_{\rm faint}\approx1.67$.  Assuming 
that this ratio holds into the mid-IR, we scaled the 2MASS data 
for EF Eri by a factor of 1.67 to match the short wavelength 
portion of the SED with the IRAC data, and repeated our modeling 
trials.  We scaled the 2MASS data instead of the IRAC data in order 
to avoid altering the observed value of the 8 $\mu$m flux density 
(which is the main target of our modeling efforts).
Modeling these scaled data yields substantially similar results to 
those described above for the unscaled data (e.g., Figure 
\ref{f:eferi_scaled} and Models 5--6 in Table \ref{t:eferi_parms}).  
The model parameters must be changed slightly from the unscaled 
versions to account for the difference in the scaled flux density 
levels; notably, the secondary star spectral type can be as early 
as L3.5 without exceeding the scaled $J$-band point (at $d=105$ pc).  
Regardless, the general trends seen in the unscaled models 
are preserved in the scaled models.

\subsubsection{V347 Pavonis}
\label{s:v347pav}

V347 Pav was discovered by \citet{pounds93} in the ROSAT Wide 
Field Camera survey. It was identified as a CV by \citet{o'donoghue93} 
and confirmed as an AM Her type by \citet{bailey95}.  
\citet{o'donoghue93} showed that the optical light curve is 
complex, with variability over the orbital cycle of approximately 
1 mag, and displaying substantial variations from one orbital 
cycle to the next. Their photometry and spectroscopy both 
suggested an orbital period of $P_{\rm orb}\sim90$ min, which 
was later refined by \citet{ramsay04} to 
$P_{\rm orb}=90.082219(52)$ min. 

\citet{bailey95} fit the optical circular polarization and light 
intensity curves with two linearly extended cyclotron emission 
regions on the surface of the WD, with a model polar field 
strength of $B=25$ MG and an orbital inclination of $i=60^{\circ}$. 
This was constrained further by \citet{ramsay96}, who used the 
orbital period\,--\,secondary star mass relation for a main 
sequence star to predict $M_{2}=0.16M_{\odot}$\footnote{We have 
used this mass in our models to predict a secondary star spectral 
type of M6.0 and radius of $R_2=0.21R_{\odot}$ \citep{cox00}.  
However, we note that this radius is substantially larger than 
the calculated Roche lobe in V347 Pav assuming 
$M_{\rm WD}=1.00M_{\odot}$; the secondary star radius of 
$R_2=0.16R_{\odot}$ for this mass from \citet{pont05} is in 
better agreement with the Roche lobe size.  In any case, the 
model SED results are not very sensitive to changes in the 
mass or radius of the stars.}, which, combined with the fact 
that they had failed to observe any eclipses of the system, 
allowed them to put an upper limit on the inclination of 
$i<78^{\circ}$.  Subsequent Stokes imaging by \citet{potter00} 
constrained the inclination more closely to $64^{\circ}<i<72^{\circ}$, 
with magnetic field strengths of $B=15$ and 20 MG for the upper 
and lower accretion poles, respectively.

\begin{figure}
\plotone{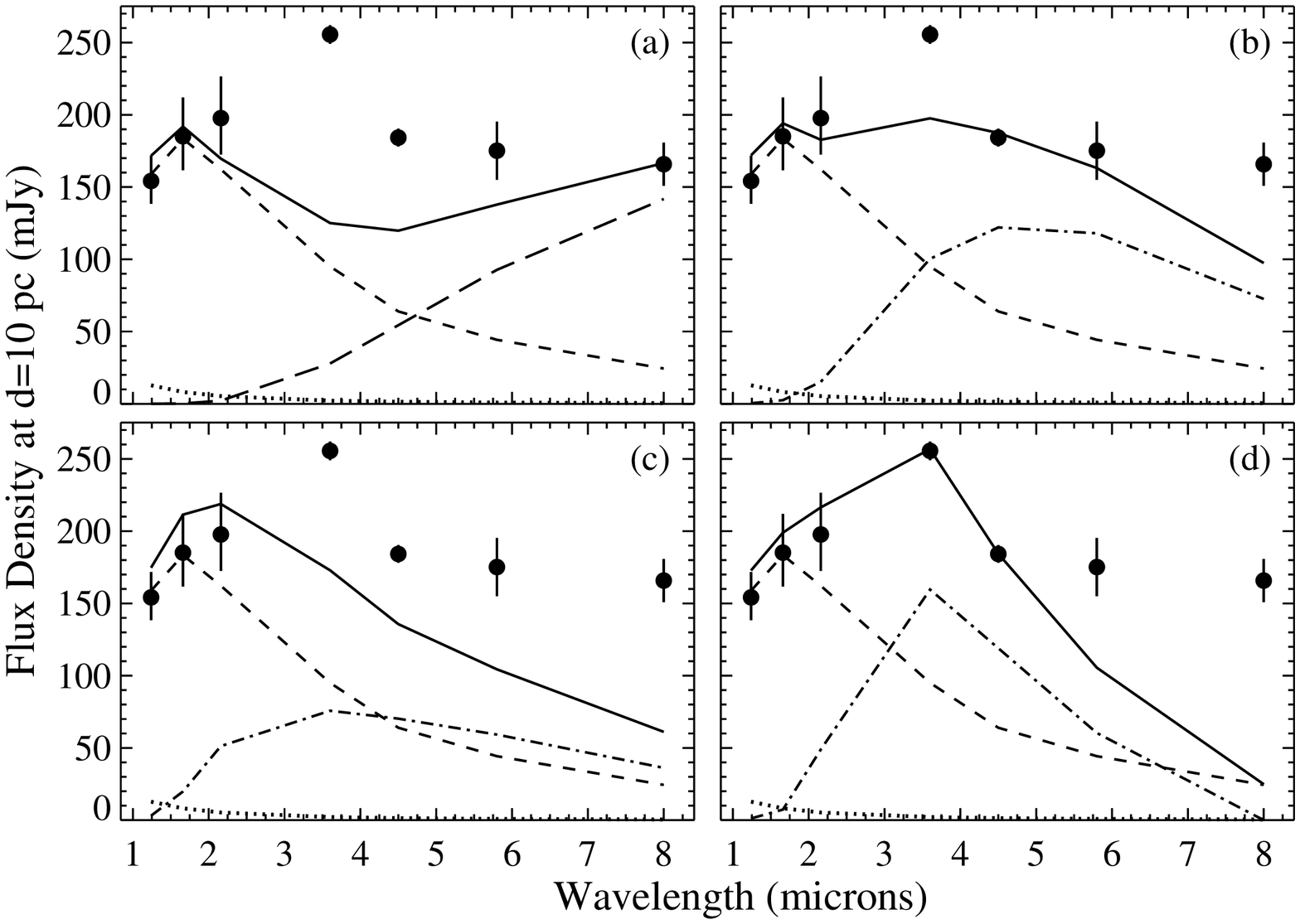}
\caption{Representative model SEDs for V347 Pav showing the observed 
data (filled circles), total model SED (solid line) and model 
components: the WD (dotted line), the secondary star (short dashed line), a circumbinary 
dust disk (long dashed line), and cyclotron emission (dot-dashed line).  
The plot panels show (a) circumbinary disk model, (b) cyclotron 
model with $m=2$ and (c) $m=3$, and (d) single-field cyclotron model 
with $m=2$.
See text and Table \ref{t:v347pav_parms} for details.
\label{f:v347pav_unscaled}}
\end{figure}

At the nominal literature distance of $d=171$ pc, the secondary 
star can be no earlier than spectral type M6.0, or the combined 
secondary star plus WD (the latter contributes less than 8\% 
of the observed SED at all wavelengths) exceeds the 
observed $J$ and $H$ points.
Figure \ref{f:v347pav_unscaled}a (Model 1 in Table \ref{t:v347pav_parms}) 
shows the model result with the addition of a circumbinary dust disk.
As with EF Eri, we can easily 
reproduce the 8 $\mu$m point, but the model underestimates 
the 3.6--5.8 $\mu$m points.  
Using a cyclotron component with $m=2$ (Figure \ref{f:v347pav_unscaled}b 
and Model 2 in Table \ref{t:v347pav_parms}), the model SED has 
the closest match to the intermediate-wavelength data, but is 
no longer bright enough at 8 $\mu$m to match the observations.  
Figure \ref{f:v347pav_unscaled}c (Model 3 in 
Table \ref{t:v347pav_parms}) shows the $m=3$ case -- larger 
values of $m$ cause the cyclotron peak to continue shifting 
toward short wavelengths until the cyclotron component appears 
as a Rayleigh-Jeans-like tail (similar in shape to the long 
wavelength end of the secondary star SED).  
Figure \ref{f:v347pav_unscaled}d (Model 4 in Table \ref{t:v347pav_parms}) 
shows an $m=2$ cyclotron component using the single-field 
approach instead of the sum-over-field-strengths.  As with EF Eri, 
this approach allows us to reproduce the sharp peak in the SED, 
which for V347 Pav is located in IRAC channel 1.  However, also 
as with EF Eri, this approach does not reproduce the 8 $\mu$m 
point, suggesting a combination of Models 1 and 4 (i.e., dust 
and single-field cyclotron) as the best approach to reproducing 
this SED.

\begin{figure}
\plotone{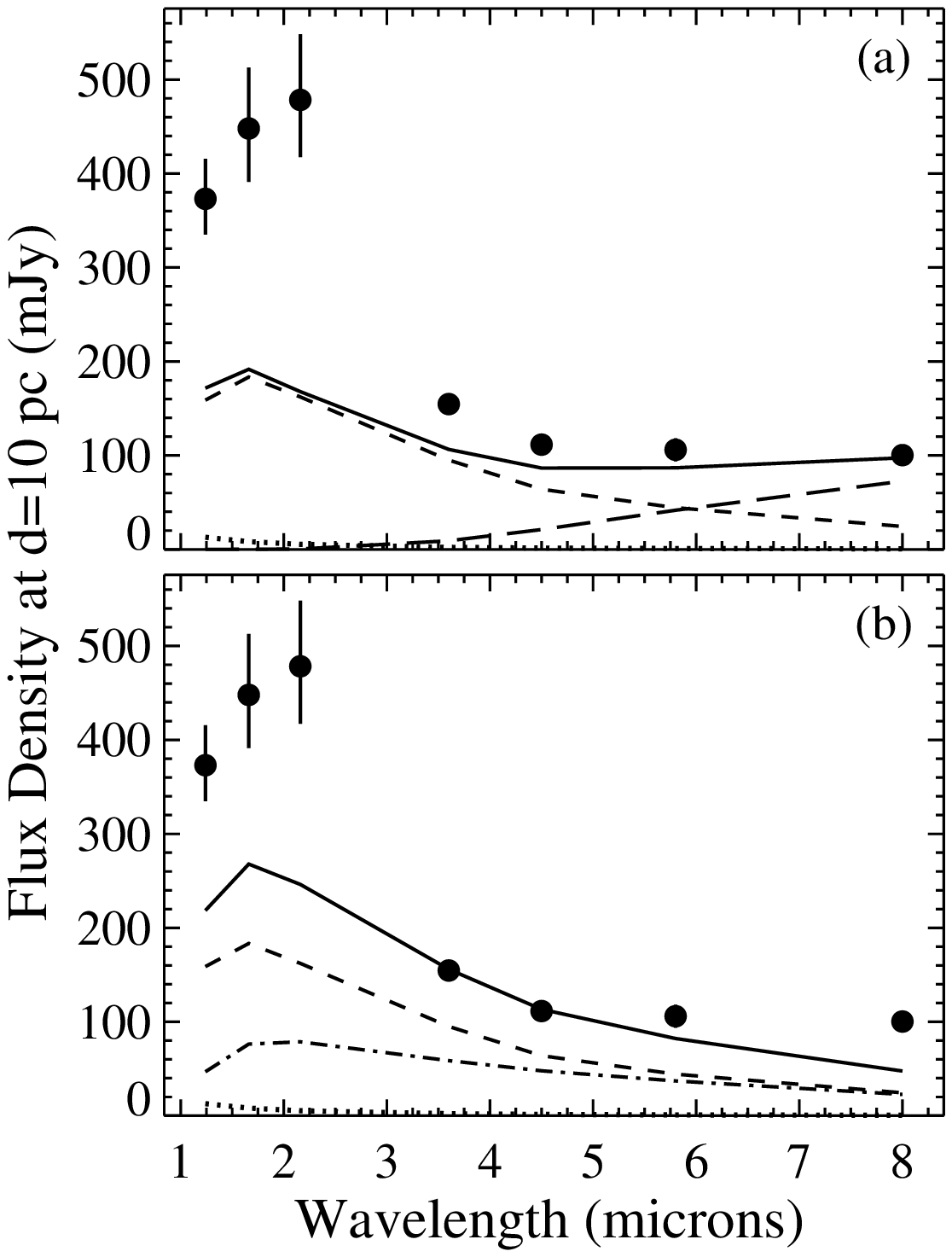}
\caption{As in Figure \ref{f:v347pav_unscaled}, but with the 2MASS 
data for V347 Pav scaled by a factor of 4 to account for potential 
orbital brightness variations.
The plot panels show (a) circumbinary disk model and (b) cyclotron 
model with $m=5$.
See text and Table \ref{t:v347pav_parms} for details.
\label{f:v347pav_scaled}}
\end{figure}

We now consider the possibility that the 2MASS and IRAC data 
were obtained at different orbital brightness states.
A high-state $H$-band photometric light curve published in 
\citet{bailey95} shows that V347 Pav is variable by a factor 
of $\approx4$ over its orbital period. 
As with EF Eri, we assume that this behavior is present in 
the mid-IR also, and attempt to correct the 2MASS and IRAC 
data sets to acccount for orbital variability.
To this end, we scaled the 2MASS flux densities by a factor of 4.0.
At the nominal distance of $d=171$ pc, the scaled SED can 
accomodate a secondary star as early as M3.5V, whose SED 
exactly matches the 3.6 and 4.5 $\mu$m points.  However, as 
noted earlier in this section, the M6.0V secondary star 
suggested by the mass estimate from \citet{ramsay96} is already 
at the limit of being too large for the calculated Roche lobe 
size, and an M3.5V star simply will not fit into the Roche lobe 
of a CV with this orbital period.  Consequently, we have 
retained the M6.0V secondary star and adjusted the distance to 
the lower limit of the \citet{araujo05} estimate, $d=133$ pc.
Figure \ref{f:v347pav_scaled}a shows the scaled data with a 
model containing a circumbinary dust disk component (Model 5 
in Table \ref{t:v347pav_parms}).  
There is considerable unmodeled flux density in the 2MASS portion 
of the SED, but even if we use an M3.5V secondary star, the model 
SED still somewhat underestimates these data points.
At all values of the transition harmonic $m\geq2$, the 
cyclotron component cannot match the 8 $\mu$m point without 
exceeding the shorter wavelength IRAC points.  
Figure \ref{f:v347pav_scaled}b (Model 6 in 
Table \ref{t:v347pav_parms}) shows a model in which the 
cyclotron component is partially optically thick ($m=2$).  
The peak of the cyclotron emission component 
shifts toward longer wavelengths, leaving a deficit of modeled 
flux density at 8 $\mu$m, but suggesting a path towards recovering 
the previously unmodeled flux at the 2MASS wavelengths.

\subsection{Class II -- Dominant Secondary Star}
\label{s:class2}

\subsubsection{VV Puppis}

VV Pup was discovered during the early 20th century \citep{vangent31} 
as a faint \cite[$V=14.5$--$18$;][]{downes01}, rapidly periodic 
($P_{\rm orb}=100.4$ min) variable star \citep{walker65}.  
Based on photometric and spectroscopic observations, \citet{herbig60} 
suggested it was a binary whose emission lines originated on 
the brighter component.  Subsequently, VV Pup was identified 
as the third known example of the AM Her class \citep{tapia77}.  

\begin{figure}
\plotone{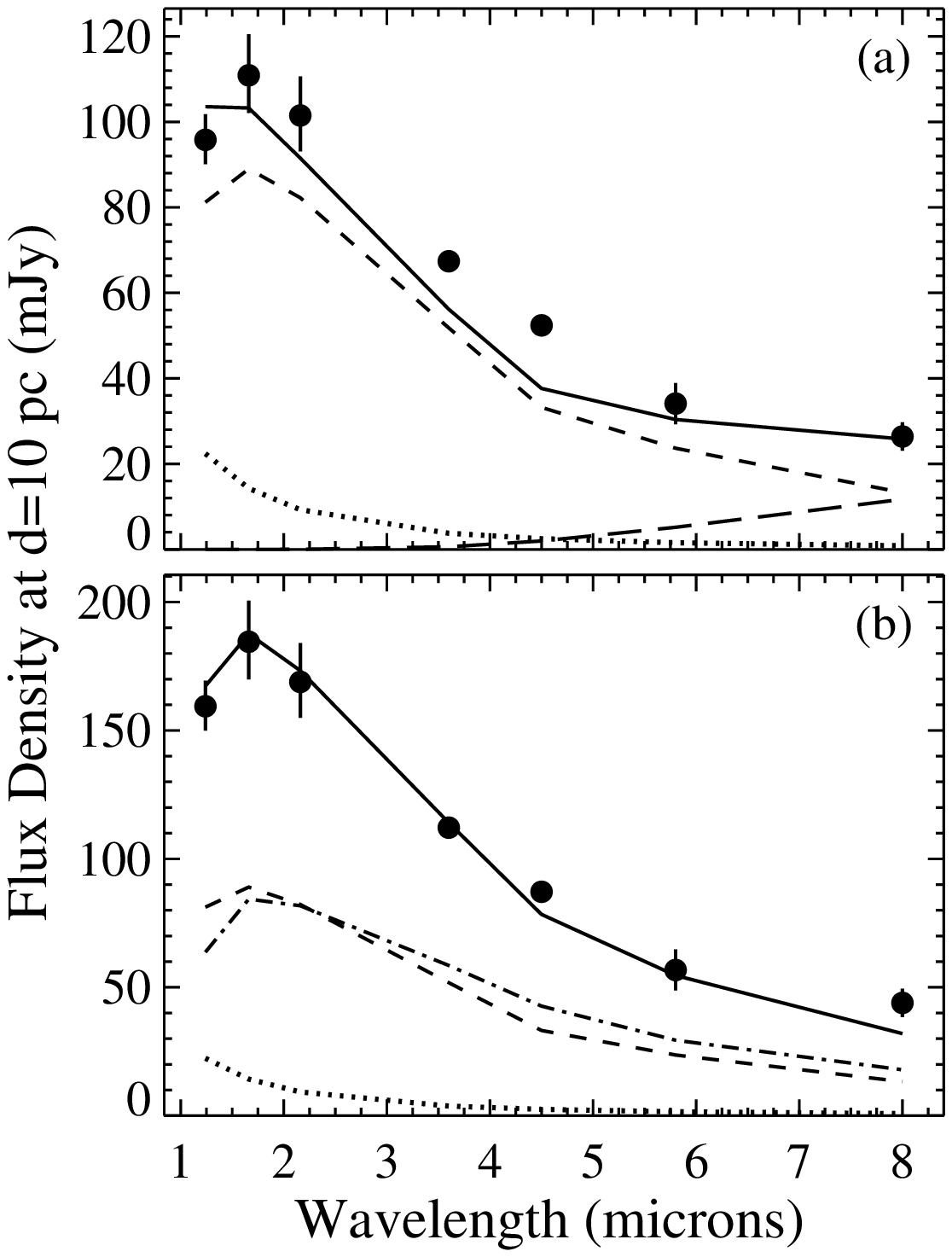}
\caption{Representative model SEDs for VV Pup showing the observed 
data (filled circles), total model SED (solid line) and model 
components:\ WD (dotted line), secondary star (short dashed line), circumbinary dust 
disk (long dashed line), and cyclotron emission (dot-dashed line).  
The plot panels show (a) circumbinary disk model and (b) cyclotron 
model with $m=3$.
See text and Table \ref{t:vvpup_parms} for details.
\label{f:vvpup_model}}
\end{figure}

VV Pup has the distinction among our sample of polars of having 
the observed SED that is most reminiscent of the falling Rayleigh-Jeans tail of a cool, main sequence star. Figure \ref{f:vvpup_model}a 
(Model 1 in Table \ref{t:vvpup_parms}) shows a model SED that 
utilizes the recent tight constraints on the WD and secondary star 
parameters provided by \citet{howell06b} with a circumbinary dust 
disk to provide the additional flux density required to match 
the observed 8 $\mu$m point. However, the model SED underestimates the observed flux densities from the H-band to 4.5$\mu$m (IRAC channel 2).  This model requires a distance 
scaling of $d=100$ pc, which is slightly more than 1$\sigma$ away from the nominal 
value of $d=129^{+29}_{-21}$ pc from the trigonometric parallax 
distance reported in \citet{howell06b}.  

For all values of the transition harmonic $m\geq2$, 
the peak of a cyclotron emission component shifts to successively shorter 
wavelengths (e.g., between the $K_{\rm s}$ and IRAC channel 1 
points for $m=2$), which causes the total model SED to exceed 
all of the shorter wavelength points when the cyclotron emission 
is scaled to fit the 8 $\mu$m point.  By moving the polar back 
to the nominal distance of $d=129$ pc and including a cyclotron 
component with $m>2$ (e.g., see Figure \ref{f:vvpup_model}b and 
Model 2 in Table \ref{t:vvpup_parms}), we can mitigate this 
problem and produce very good model SEDs (with better agreement 
at 3.6 and 4.5$\mu$m and only a small flux density deficit at 8$\mu$m).  
However, this form of the model is not very discriminatory, 
as the shape of the cyclotron component SED for $m>2$ is very 
similar to that of the secondary star, so this model is 
basically equivalent to including only the stellar components 
and simply scaling the distance up or down until the best match 
to the observed data is obtained.

\subsection{Class III -- Bright 2MASS, Flat IRAC}

We suspect that the polars in Class III were in high accretion 
states during the 2MASS observations due to the shape of the 
bright 2MASS portion of their SEDs, which could be indicative 
of the falling Rayleigh-Jeans-like tail of accretion-generated 
luminosity. If present, then this (unmodeled) component likely 
contributes at 3.6 and 4.5 $\mu$m in addition to dominating the 
2MASS portion of the SED.  
We have not attempted to model this additional component because 
the exact characteristics of the accretion-generated luminosity 
are not known in detail, and would depend on a large number of 
(largely unknown or poorly constrained) additional parameters 
(e.g., mass transfer rate, plasma flow and magnetic field 
geometry, accretion spot size and shape, etc.).  The additional 
complexity that would be introduced by adding another model 
component is not warranted for our sparsely sampled SEDs.

\subsubsection{V834 Centauri}

\citet{mason83} identified a 15th magnitude, blue, emission 
line (\ion{H}{1}, \ion{He}{1,II}) star (subsequently named V834 Cen) 
as the optical counterpart of the variable X-ray source 
H1409-45/E1405-451 \citep{jensen82}.  Their time-resolved optical 
photometry revealed a period of $P_{\rm orb}=101.52(2)$~minutes 
with an amplitude of $\sim1$~mag.  In addition, rapid variability 
on timescales of 1--3~seconds was observed; this was later shown 
to be quasiperiodic in nature, with a coherence time 
of $\sim 1$~min \citep{larsson85,larsson92,imamura00}.  
\citet{wright88} detected V834 Cen in the radio, at a peak level 
of 35~mJy at 8.4~GHz.  The radio flux showed variability on 
timescales as short as 1 min.  Based on these characteristics 
(coupled with its X-ray emission, and circular and linear 
polarization observed by \citealt{tapia82} and 
\citealt{visvanathan83}), \citet{mason83} suggested that this 
object was an AM Her type CV.

\begin{figure}
\plotone{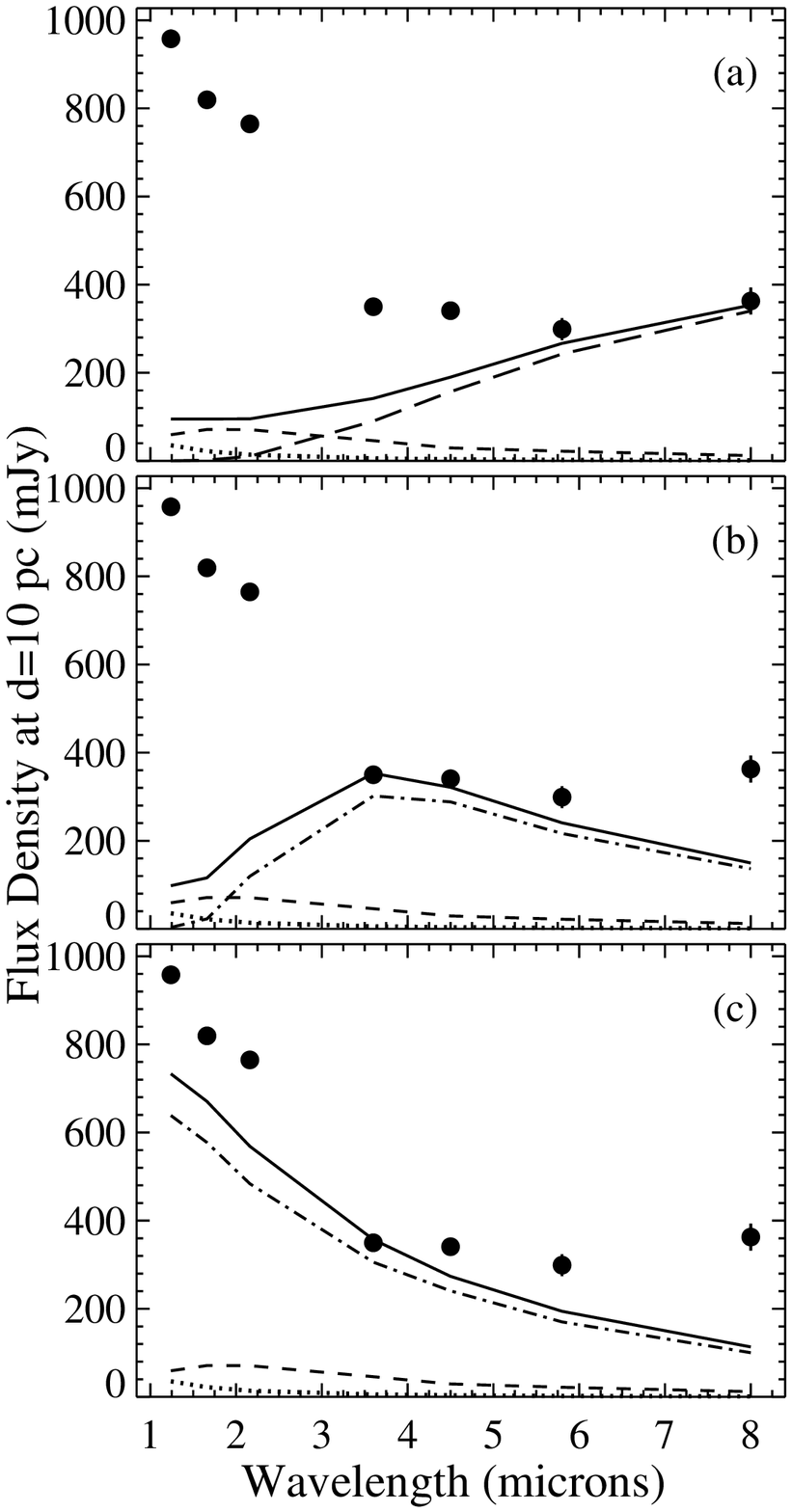}
\caption{Representative model SEDs for V834 Cen showing the 
observed data (filled circles), total model SED (solid line) and 
model components:\ WD (dotted line), secondary star (short dashed line), circumbinary 
dust disk (long dashed line), and cyclotron emission (dot-dashed 
line). The plot panels show (a) circumbinary disk model and (b) cyclotron 
model with $m=2$ and (c) $m=6$.
See text and Table \ref{t:v834cen_parms} for details.
\label{f:v834cen_model}}
\end{figure}

Recently, \citet{mennickent04} have matched the secondary star 
features in near-infrared (1--2.5 $\mu$m) spectra of V834 Cen 
with a spectral type of M8$\pm$0.5V.
An M8V star is quite faint, so we have used a minimum distance 
of $d=121$ pc in our models \citep{araujo05}.  
Figure \ref{f:v834cen_model}a (see Model 1 in 
Table \ref{t:v834cen_parms}) shows the model SED with a 
circumbinary dust disk, which reproduces the observed 8 $\mu$m 
point but underestimates the flux densities at shorter wavelengths.  
The secondary star would contribute significantly to the total model 
SED only at much smaller distances ($d\approx50$ pc) than estimated 
in the literature.  
Figure \ref{f:v834cen_model}b (Model 2 in Table \ref{t:v834cen_parms}) 
shows the model with the circumbinary disk replaced by 
cyclotron emission with $m=2$.
The cyclotron profile peak is centered around 3.6--4.5 $\mu$m, and 
even when the cyclotron emission is scaled up so that all of the 
3.6 $\mu$m flux density is accounted for, it still underestimates 
the 8 $\mu$m point.  Figure \ref{f:v834cen_model}c (Model 3 in 
Table \ref{t:v834cen_parms}) shows a largely optically thick ($m=6$) 
cyclotron component.  In this case, the cyclotron component can 
account for some of the 
flux density in the bright 2MASS portion of the SED, but still 
cannot reproduce the 8 $\mu$m point.

\subsubsection{GG Leonis}

GG Leo was discovered by the ROSAT All Sky Survey and subsequently 
identified as a polar by \citet{burwitz98}. 
They derived the orbital period $P_{\rm orb}$ = 79.87946(7) min 
from X-ray dip timings and concluded that the shape of the light 
curve is consistent with accretion at only one of the WD poles. 
The optical light curve presented in \citet{burwitz98} is variable by 
about 1 mag over the orbital period, and their photopolarimetry 
confirmed the one-pole accretion model and allowed them to derive 
a magnetic field strength of $B\sim23$ MG in the accretion region.  
A fit to the low-state spectrum and the cyclotron emission yielded 
WD parameters of $0.8M_{\odot} < M_{\rm WD} < 1.2M_{\odot}$ and 
$T_{\rm WD}=8000$ K, although the authors note that temperatures 
up to $T_{\rm WD}=12,000$ K cannot be excluded. 
\citet{ramsay04} give a revised mass estimate for the WD of 
$M_{\rm WD} = 1.13\pm$0.03~M$_{\odot}$ based on X-ray data. 

\citet{burwitz98} estimated the secondary star parameters from 
the orbital period and the radial velocity profile, adopting a 
radius of $R_{2}\sim0.13R_{\odot}$, a mass of $M_2\sim0.09M_{\odot}$, 
and a spectral type of M5V or later.  (We have used a spectal type 
of M7.5V, which corresponds to these mass and radius values, in 
our models.)   They then used the non-detection of spectral features 
expected from the secondary star to put a lower limit on the distance 
of $d=100$ pc, while the upper distance limit of $d=300$ pc is based 
on the fact that any larger distances would require unrealistically 
high accretion rates. 
Unless the distance to GG Leo is substantially larger than 300 pc, 
the secondary star spectral type cannot be earlier than M4V or the 
model flux at 3.6 $\mu$m will exceed the observed value.  
In addition, {\em no} secondary star template from the Patten et al. 
sample (even at early spectral types) can reproduce the observed 
2MASS SED without greatly exceeding the observed IRAC data.
At a median distance of $d=200$ pc, the M7.5V secondary star SED 
underestimates the 3.6 and 4.5 $\mu$m points by about 50\%, leaving 
room for the unmodeled accretion-luminosity component.

\begin{figure}
\plotone{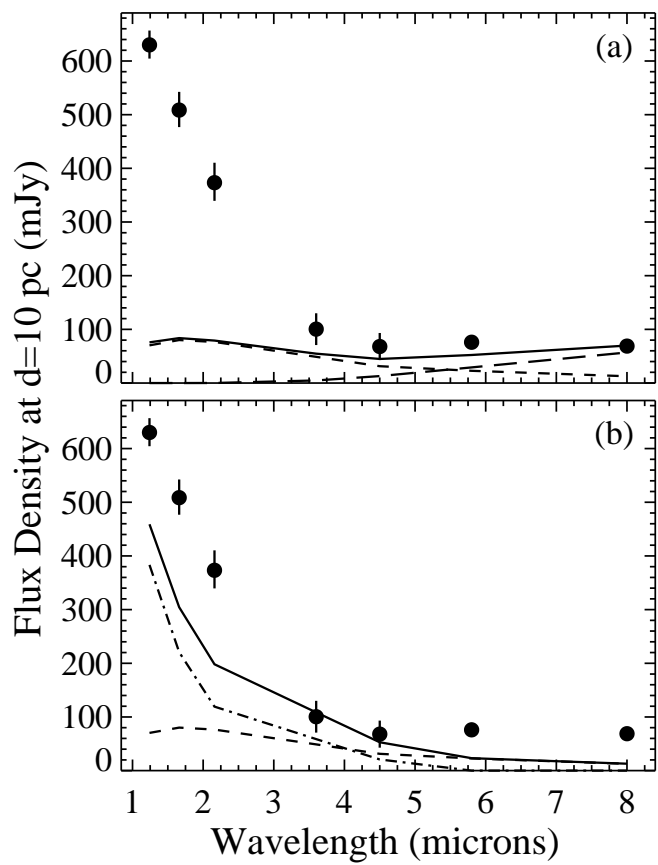}
\caption{Representative model SEDs for GG Leo showing the observed 
data (filled circles), total model SED (solid line) and model 
components:\ secondary star (short dashed line), circumbinary dust 
disk (long dashed line), and cyclotron emission (dot-dashed line).  
At maximum, the WD contributes less than 1\% of the total observed 
flux density and is not plotted.  
The plot panels show (a) circumbinary disk model and (b) single-field 
cyclotron model with $m=10$.
See text and Table \ref{t:ggleo_parms} for details.
\label{f:ggleo_model}}
\end{figure}

As shown in Figure \ref{f:ggleo_model}a (Model 1 in 
Table \ref{t:ggleo_parms}), a circumbinary dust disk can reproduce 
the 8 $\mu$m point.  A similar result can be obtained even if we 
assume the minimum distance ($d=140$ pc) at which the M7.5V 
secondary star exactly matches the 3.6 and 4.5 $\mu$m points.  
For all values of the transition harmonic $m\geq2$, the peak of the 
cyclotron component is located at short wavelengths, making it 
impossible to reproduce the 8 $\mu$m point without greatly 
exceeding the 3.6--5.8 $\mu$m points.  If we use only the 
single-field cyclotron prescription, and assume completely 
optically thick emission ($m=10$), 
then the resultant model trends toward reproducing the bright 
2MASS portion of the SED -- suggesting that cyclotron emission, 
instead of or in addition to, accretion luminosity could also be 
responsible for the 2MASS data -- but still cannot reproduce 
the 8 $\mu$m point (see Figure \ref{f:ggleo_model}b and Model 2 
in Table \ref{t:ggleo_parms}).

\subsubsection{MR Serpentis}

MR Ser was initially discovered in the Palomar-Green survey for 
UV excess objects \citep{green82}; it turned out to be the only 
magnetic CV discovered in the PG survey.  \citet{liebert82} 
performed the first detailed observational study of MR Ser, 
and measured an orbital period of $P_{\rm orb}=113.56$ min 
from polarimetric observations.  This was later refined by 
\citet{schwope91}, who used time-resolved spectroscopic 
observations of a narrow emission line component (believed 
to originate on the secondary star) to measure 
$P_{\rm orb}=113.47$ min (also see \citealt{schwope93}).
\citet{wickramasinghe91} observed shallow cyclotron humps 
in their optical spectra, which varied with phase in amplitude 
and wavelength.  \citet{harrison05} found cyclotron emission 
from the $B=26$ MG WD magnetic field contaminating the blue 
end of their $K$-band spectrum, but the red end of the spectrum 
was not significantly affected by cyclotron emission.

MR Ser dropped into a low state in 1985, at which time the 
optical spectra showed weak H--alpha emission and absorption 
features of an M dwarf secondary star \citep{mukai85}. 
\citet{szkody88} found that even in the low state, H and He 
emission lines were still present, indicating that there was 
still some ongoing accretion. \citet{schwope93} identified 
features of the secondary star in long wavelength optical 
spectra of MR Ser, corresponding to a spectral type of M5--6V 
(also see \citealt{mukai86}).  
However, \citet{harrison05} estimated a somewhat later 
secondary star spectral type, M8V, from an infrared spectrum.  

\begin{figure}
\plotone{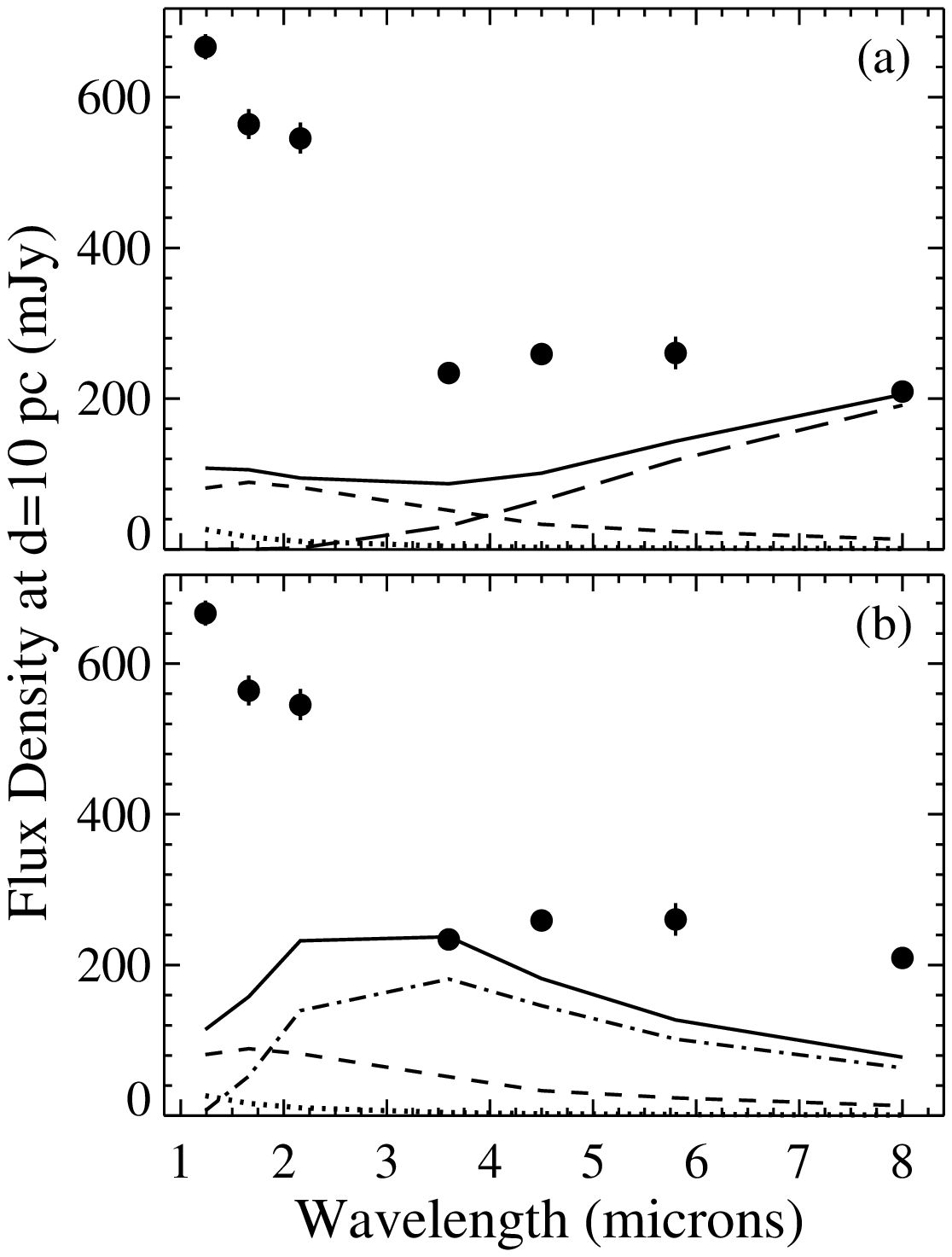}
\caption{Representative model SEDs for MR Ser showing the observed 
data (filled circles), total model SED (solid line) and model 
components:\ WD (dotted line), secondary star (short dashed line), circumbinary dust 
disk (long dashed line), and cyclotron emission (dot-dashed line).    
The plot panels show (a) circumbinary disk model and (b) cyclotron 
model with $m=2$.
See text and Table \ref{t:mrser_parms} for details.
\label{f:mrser_model}}
\end{figure}

Of the three polars in our Class III, MR Ser and V834 Cen have 
the most similar observed SEDs.  It is no surprise, then, that 
our modeling results for MR Ser are very similar to those for 
V834 Cen.  We utilized a minimum distance of $d=134$ pc (see 
Table \ref{t:parameters}) in order to maximize the contribution 
from the faint secondary star.
At this distance, the spectral type of the secondary star cannot 
be earlier than about M5.5V, or the model 3.6 $\mu$m point will 
be brighter than observed.
Figure \ref{f:mrser_model}a (Model 1 in Table \ref{t:mrser_parms}) 
shows the model for MR Ser containing an M7V secondary star 
(consistent with the mass estimate of $M_2=0.1M_{\odot}$ -- see 
Table \ref{t:parameters}) and a circumbinary dust disk.  
As with the similar model for V834 Cen, this reproduces the 
observed 8 $\mu$m flux density, but underestimates the flux 
density at all shorter wavelengths.  
As also found for V834 Cen, at values 
of $m\geq2$, the cyclotron profile peak shifts toward shorter 
wavelengths, making it impossible to reproduce the 8 $\mu$m point 
without exceeding one of the shorter wavelength IRAC points 
(e.g., see Figure \ref{f:mrser_model}b and Model 2 in Table 
\ref{t:mrser_parms}).  

An interesting aspect of the observed IRAC SED for MR Ser is 
that the 4.5 and 5.8 $\mu$m points are brighter than the 3.6 
and 8.0 $\mu$m points.  This behavior is not seen in any of 
the other polars in our sample (see Figure \ref{f:seds}).  
Reproducing this shape is not possible using only circumbinary 
disk and/or cyclotron components, without invoking an 
additional source of flux that peaks between IRAC channels 
2 and 3; for example a blackbody component with $T\approx800$ K 
(also see Appendix \ref{s:csd}).

\subsection{Sample ``Best'' Models}
\label{s:best}

While keeping in mind our warning about the (in)ability of 
our data to constrain complex models, we also recognize that 
it is somewhat unsatisfying to just consider models that are 
designed to only reproduce the observed 8 $\mu$m point with 
little regard to the SED shape at shorter wavelengths (other 
than to avoid exceeding the observed flux densities).
To that end, we will present a sample ``best'' model for each 
of our polars (see Table \ref{t:best_parms}).  
These models are not constrained to be unique solutions and 
should not be considered as definitive physical representations 
of the polars.
They simply illustrate that it is possible to create reasonable 
reproductions of the observed IR SEDs for our polars, which, 
given additional data, could be developed into much more 
physically detailed models.
However, the models shown in this section do not demand that 
the observed SEDs must be fit using only the model components 
utilized here.

\begin{figure}
\plotone{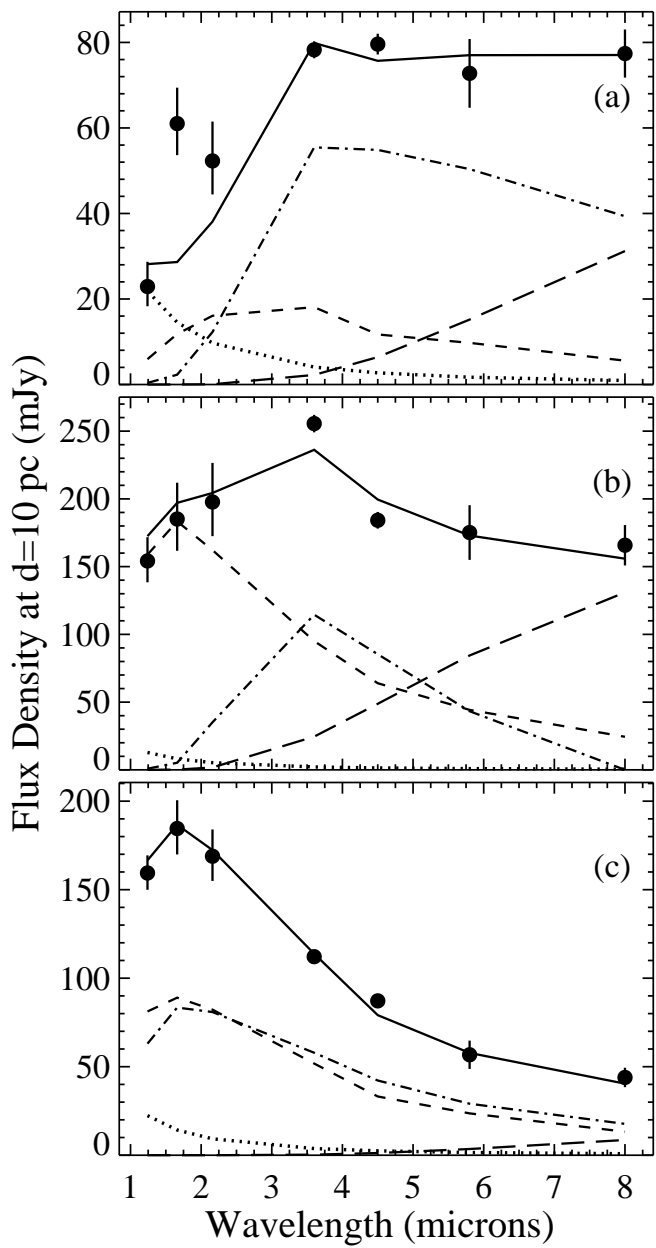}
\caption{Sample model SEDs that best reproduce the observed data 
for (a) EF Eri (unscaled), (b) V347 Pav, and (c) VV Pup,
showing the observed data (filled circles), total model SED (solid 
line) and model components:\ WD (dotted line), secondary star (short dashed line), 
circumbinary dust disk (long dashed line), and cyclotron emission 
(dot-dashed line).  
See text and Table \ref{t:best_parms} for details.
\label{f:best1}}
\end{figure}

\begin{figure}
\plotone{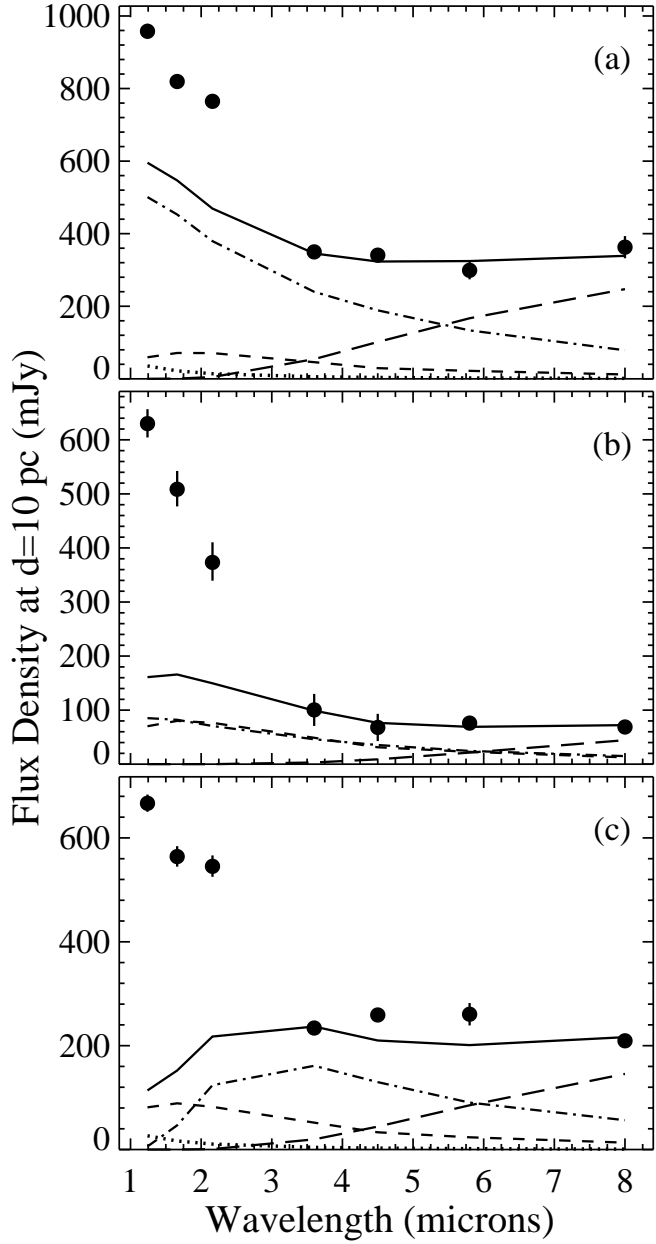}
\caption{As in Figure \ref{f:best1}, but showing sample ``best'' 
models for 
(a) V834 Cen, (b) GG Leo, and (c) MR Ser.
See text and Table \ref{t:best_parms} for details.
\label{f:best2}}
\end{figure}

In all cases the ``best'' models involve combining the stellar 
components with cyclotron {\em and} circumbinary disk emission.  
Typically, the circumbinary disk emission dominates in producing 
the excess 8 $\mu$m emission, whereas the cyclotron emission 
dominates at shorter wavelengths.
Figure \ref{f:best1} shows sample ``best'' models for EF Eri 
(with unscaled 2MASS data), V347 Pav (unscaled), and VV Pup. In the case of EF Eri, we have optimised the model to best reproduce the IRAC SED, while neglecting the bright 2MASS H-band point (see section 4.1.1 for an example of fitting the H-band point).
Figure \ref{f:best2} shows sample ``best'' models for V834 Cen, 
GG Leo, and MR Ser.  In these cases, we have not attempted to 
reproduce the bright 2MASS portions of the SED.  We can generally 
achieve good reproductions of the IRAC portions of the SEDs.  
Of course, there is likely contamination from the 
accretion-generate luminosity in IRAC channels 1 and 2, which, 
if modeled, would cause these ``best'' models to overestimate 
the flux density at those wavelengths.  However, this can easily 
be accomodated by making small adjustments to the distance, 
circumbinary disk inner temperature, and/or cyclotron emission 
strength.  Out of all six of our polars, the ``best'' model for 
VV Pup comes closest to exactly reproducing the observed SED 
(see \S\ref{s:class2}).

\section{Discussion and Conclusions}
\label{s:discussion}

We emphasize again that the results of our modeling should be 
considered less as specific, detailed pictures of each individual 
polar, and more as an illustration of the general mid-IR properties 
of the class of polars as a whole.  With that in mind, we can 
describe the general trends that our modeling efforts have revealed:

(1) There is excess flux density at 8 $\mu$m above that 
expected from just the stellar components in all of our target 
polars.  Our capability to obtain sensitive IR observations 
has improved dramatically in recent years, especially following 
the launch of the Spitzer Space Telescope.  Evidence for IR 
excess in isolated WDs is becoming increasingly common, and 
has been convincingly linked to the presence of circumstellar 
dust disks.  Now, CVs and LMXBs have joined the pantheon of 
systems containing compact objects that show excess luminosity at 
long wavelengths.
Among the CVs, polars (which lack the bright accretion disk of non-magnetic
systems) may offer the best opportunity to detect the comparatively faint
signature of circumbinary dust.  The likelihood of detecting the dust
emission is also improved by observing the CVs (polars or otherwise) 
with the shortest
orbital periods (i.e., $P_{\rm orb}\lesssim90$ min), which contain the
lowest mass (hence, least luminous) secondary stars.

(2) In all cases, a circumbinary dust disk can reproduce the 
observed 8 $\mu$m flux density.  
Equivalent results can be obtained using either optically thick 
or optically thin disks.  Although we have only presented model 
SED results from the computationally simpler optically thick 
disks here, we have used equivalent models containing optically 
thin disks to provide an estimate of the masses of dust required 
to explain the observed mid-IR excesses in our polars.
Typical values for the total dust mass in our optically thin 
circumbinary disk models are $M_{\rm dust}\sim10^{15}$--$10^{17}$ g, 
or about $10^{-19}$--$10^{-17}M_{\odot}$.  This is only a very 
small fraction of the annual mass transfer budget in 
these short-orbital-period CVs, 
at typical rates of $\dot{M}\lesssim10^{-11}M_{\odot}$ yr$^{-1}$ 
\citep{howell95}.  It is also smaller than the $10^{20}$--$10^{22}$ g 
of dust produced during a superoutburst of a short-orbital-period 
CV \citep{ciardi06}.  These circumbinary disks 
could be created (and, if necessary, replenished) from only a 
comparative trickle of outflowing matter over the course of 
millions or billions of years.

In addition, the total dust mass in our model circumbinary disks 
is many orders of magnitude lower than the $\sim10^{29}$ g of 
material that \citet{taam03} require to significantly affect CV 
evolution.  \citet{muno06} 
also found a paucity of circumbinary material in their LMXB 
systems ($M_{\rm dust}\sim10^{22}$ g), as did \citet{dubus04} 
from their circumbinary disk model calculations for CVs
($M_{\rm dust}\lesssim10^{24}$ g).
As noted by 
\citet{muno06}, even if the gas-to-dust mass ratio in the 
circumbinary disk is 100:1, then the total disk mass is still 
several orders of magnitude too small to satisfy the \citet{taam03} 
condition.  If our estimates of the circumbinary disk masses are 
correct, then either they cannot be responsible for the observed 
discrepancies in CV angular momentum loss rates as suggested by 
\citet{taam03}, or the interaction between the circumbinary disk 
and the inner binary system is poorly understood. 

(3) On the other hand, a completely optically thin cyclotron 
emission model component typically has a calculated SED shape 
that is indistinguishable 
from, or very similar to, that of the circumbinary dust disk.  
Hence, we cannot exclude the specific case of cyclotron emission 
with transition harmonic $m=1$ as the source of the excess mid-IR 
flux density.  However, when the model cyclotron emission is completely 
optically thin, it also does not contribute strongly at shorter 
wavelengths and the resultant model SEDs underestimate the flux 
density at short wavelengths.  
In addition, observational evidence for completely optically
thin cyclotron emission has not yet been found in {\em any} polar.
Since the circumbinary disk emission 
always contributes most strongly at 8 $\mu$m (with little or no 
flux density contriubution shortward of 3.6 $\mu$m), it seems more 
likely that cyclotron emission in these polars is not completely 
optically thin, and can account for some of the shorter wavelength 
flux density (e.g., see the ``best'' models discussed in \S\ref{s:best}).

(4) In all but one case, cyclotron emission with 
$m\geq2$ is too faint at 8 $\mu$m to reproduce the observed data 
when it is scaled so that it does not exceed any of the shorter 
wavelength IRAC points.  The exception to this is VV Pup, which 
has the smallest mid-IR excess compared to the model stellar 
components.  For this polar, optically thick cyclotron emission 
in combination with the stellar components can adequately reproduce 
the observed SED.  However, this result for VV Pup does not really 
imply a preference for cyclotron emission since the optically 
thick cyclotron SED mimics the shape of the falling Rayleigh-Jeans 
tail of the secondary star SED.  Hence, this model devolves to the 
equivalent of using only the secondary star model SED and scaling 
the distance to match the observed data.
There are several possible resaons why VV Pup could contain the 
apparently smallest mid-IR excess of our sample of polars.  
It is probably {\em not} due to masking of the dust signature
by a bright secondary star, since
VV Pup does not have the longest orbital period of our sample
(and, hence, does not have a correspondingly much earlier 
spectral type secondary star than the other polars -- both 
MR Ser and V347 Pup have somewhat longer orbital periods and 
presumably similar secondary stars to VV Pup).
If the IR excess is caused largely or solely by dust, then VV Pup
possibly just implies that the amount or emissive characteristics
of dust is not the same in all polars.  The most likely explanation, 
however, is that VV Pup has the highest estimated inclination of 
our sample of polars, so the projected radiating area of the 
circumbinary disk is smallest.

In summary, while a circumbinary dust disk can always explain the 
observed 8 $\mu$m excess in our target polars, only the specific 
case of completely optically thin cyclotron emission (which mimics 
the SED shape of a circumbinary disk but has not yet been observed
in any polar) can also independently 
explain the mid-IR excess.  However, even in this special case, the cyclotron 
emission can not be used to account for excess flux density at 
shorter wavelengths (where any feasible circumbinary disk emission 
is also negligible).  In the presence of cyclotron emission that 
is at all optically thick (and, hence, peaks at shorter IR 
wavelengths), circumbinary dust disk emission would still have 
to be present in order to explain the 8 $\mu$m excess in our 
polars.  If the mid-IR excess in our polars {\em is} produced, 
either solely or in part, by dust in circumbinary disks, then 
the implied mass of dust is too small by many orders of magnitude 
to affect CV evolution as described in \citet{taam03}.

\acknowledgements
This work is based in part on observations made with the Spitzer 
Space Telescope, which is operated by the Jet Propulsion Laboratory, 
California Institute of Technology, under a contract with NASA.  
Support for this work was provided by NASA through an award issued 
by JPL/Caltech.  This publication makes use of data products from 
the Two Micron All Sky Survey, which is a joint project of the 
University of Massachusetts and the Infrared Processing and 
Analysis Center/Caltech, funded by NASA and the National Science 
Foundation.  We acknowledge with thanks the variable star 
observations from the AAVSO International Database contributed 
by observers worldwide and used in this research. Last, but by no means least, we would like to thank the anonymous referee for the thoughtful and thorough review of our paper.

\appendix
\section{On the Possible Presence of a White Dwarf Circumstellar Dust Disk in Polars}
\label{s:csd}

The SED shape produced by a single-temperature blackbody component 
can also be produced by an optically thick 
\emph{circumstellar} dust disk around the WD in a polar.
Such a disk is 
exactly analogous to the WD debris disks described in \citet{jura03} and 
\citet{becklin05}, which were used as the basis for our consideration of 
the circumbinary disks.  
While the presence of circumstellar disks around isolated WDs is 
becoming increasingly ``normal'' as detailed IR observations of these 
objects are acquired, the existence of a WD circumstellar disk in a 
CV would be a rather novel occurence.  
Nonetheless, we will briefly explore the nature and plausibility of 
such a system component (although further exploration of its 
characteristics -- and even its existence -- will demand more detailed 
mid-IR observations than are presently available).

The circumstellar disk is centered around the WD, and extends from the 
sublimation radius close to the WD (i.e., the radius at which the disk 
temperature, according to equation \ref{e:disktemp}, is in the range 
1000--2000 K, typically $\sim$5--10$R_{\rm WD}$), out to the tidal 
truncation radius of 0.49 times the binary separation ($\approx$80\% 
of the WD Roche lobe radius; e.g., \citealt{papaloizou77}).  
The overall hotter temperature of this circumstellar disk can produce 
emission at the short IRAC wavelengths with an SED shape comparable to 
a single-temperature blackbody component, 
while its smaller radiating area (compared to a circumbinary disk) does 
not require moving the CV to large distances in order to avoid exceeding 
the observed flux density levels.  

It is important to clarify the nature of this circumstellar dust disk.  
It is more analogous to a passive planetary ring system (\`{a} la Saturn) 
than to a normal CV accretion disk.  For example, it is assumed to be a 
uniform disk that is geometrically thin, with a typical thickness 
comparable to the size of the dust grains.  Collisional interactions 
between dust grains are assumed to be relatively rare (compared to a 
viscous accretion disk).  Consequently, we assume that there is no 
(or only negligible) accretion-generated luminosity in the circumstellar 
dust disk.  
Several potential sources of dust in a CV are listed in \S\ref{s:cbd}.  
We now add two more possibilities that apply particularly to dust in 
a circumstellar disk around the WD:\ 
(1) a tidally disrupted asteroid or comet, as described in \citet{jura03} 
and \citet{becklin05}, and 
(2) dust produced in the outer atmosphere or wind of the secondary star 
\citep{lunine89, allard01, leggett02, tsuji02} and transferred into 
the Roche lobe of the WD through the L1 point.

In the polars, the WD magnetic field prevents the formation of an 
accretion disk because the plasma emerging from the L1 point is 
captured onto magnetic field lines and funneled directly to the 
poles of the WD.  
This mechanism should not work efficiently to prevent 
the formation of a dust disk that is composed of un-ionized (or
otherwise uncharged) material that will not interact strongly with 
the magnetic field. 
However, dust grains in space that are exposed to UV radiation will 
generate a flux of photoelectrons and develop a positive surface 
charge that rapidly reaches an equilibrium value \citep{horanyi96}.  
The amount of equilibrium charge has some dependence on the grain 
composition, and the grain size determines the timescale over which 
the equilibrium charge is reached (with larger grains reaching the 
equilibrium charge faster).  
We define the critical surface charge, $Q_{\rm crit}$, on a dust 
grain such that,   
for larger charges, the dust will interact with the WD magnetic 
field in a polar and formation of a dust disk would be impeded in 
much the same way as for a normal CV accretion disk.  
For smaller charges, the dust and magnetic field will not interact 
appreciably, and the circumstellar dust disk can form even around 
the magnetic WD in a polar.

By comparing the gravitational and magnetic forces acting on a dust 
grain emerging from L1 and undergoing free-fall acceleration toward 
the WD, we can estimate the value of $Q_{\rm crit}$.  
The gravitational force, $F_{\rm G}$, is given by

\begin{equation}
F_{\rm G} = \frac{GmM_{\rm WD}}{r^{2}},
\end{equation}

\noindent where $m$ is the mass of a dust grain, $M_{\rm WD}$ is the WD mass, 
$r$ is the distance from the center of the WD to the dust grain, 
and $G$ is the gravitation constant.  
The magnetic force acting on the dust grain, $F_{\rm m}$, is given by

\begin{equation}
F_{\rm m} = Q v B_{\rm WD},
\end{equation}

\noindent where $Q$ is the surface charge on a grain of dust, $v$ is the 
free-fall velocity of the dust (itself a function of distance from 
the WD), and $B_{\rm WD}$ is the WD surface magnetic field 
(magnetic flux density).  After equating these two forces, 
we can solve for $Q_{\rm crit}$ to obtain

\begin{equation}
Q_{\rm crit} = \frac{\left(\frac{1}{2} G M_{\rm WD}\right)^{1/2} m 
R_{\rm sub}^{3/2}}{R_{\rm WD}^{3} B_{\rm WD}},
\end{equation}

\noindent where we have considered the maximum potential interaction between 
the WD magnetic field and the dust by solving for $Q_{\rm crit}$ 
as close as possible to the WD, at the dust sublimation radius, $R_{\rm sub}$.

We have calculated values of $Q_{\rm crit}$ using the system parameters 
of MR Ser, and assuming $R_{\rm sub}=7R_{\rm WD}$, which corresponds to 
$T_{\rm sub}=1500$ K (the midpoint of the possible 
range of sublimation temperatures).  
For spherical dust grains with uniform density of 3 g cm$^{-3}$ and 
radii of 1, 10, and 100 $\mu$m, we obtain 
$Q_{\rm crit} \approx 1.9\times10^{n}e$, where $n$ = 2, 5, and 8, 
respectively \footnote[1]{We note in passing that our calculated values of $Q_{\rm crit}$ are several orders of magnitude smaller than the corresponding values, $Q_{\rm max}$, at which the electrostatic tensile stress in the dust grain interiors would be sufficient to fracture (i.e., destroy) the grains \citep[e.g., following the calculation in][]{draine79}.}.  
\citet{kempf04} measured the charge on in-situ interplanetary dust 
grains in the solar system using the Cosmic Dust Analyzer on the 
{\em Cassini} spacecraft.  
They found values of $\sim$8--34$\times10^{3}e$ for $\sim10$ $\mu$m, 
spheroidal grains.  
This is $\sim$5--25 times smaller than our $Q_{\rm crit}$ for 
similarly sized grains.
\citet{sickafoose00} have experimentally investigated the 
photoelectric charging of dust grains, and find an equilibrium surface 
charge of $\approx4\times10^{4}e$ for graphite grains with radii of 
$\approx50$ $\mu$m in the presence of UV illumination with a minimum 
wavelength of $\approx2000$\AA.  
Our value of $Q_{\rm crit}$ for a grain of this size is 
$\approx2\times10^{7}e$, 500 times larger than the charge 
measured by \citet{sickafoose00}.  
We infer that the equilibrium charge that is likely to form 
on dust grains in WD circumstellar disks in our polars is 
substantially smaller than the charge required to produce significant 
interaction between the dust and the WD magnetic field at distances 
from the WD larger than the sublimation radius.
Thus, the presence of a circumstellar dust disk around the WD in a 
polar, while possibly improbable, is not an impossible scenario.

This calculation was motivated by the fact that, during our modeling process, we noticed that the unusual 
shape of the IRAC SED of MR Ser matches very closely to that 
expected from only a WD circumstellar disk.  If the secondary 
star in MR Ser actually contributes very little to the total SED 
(or at least contributes more or less equally in all IRAC channels), 
then the observed mid-IR SED can be reproduced very well through 
the addition of only a WD circumstellar disk with an inner 
temperature of 1300 K.  However, in this case, the secondary star 
must be exceptionally faint in order to prevent the 3.6 $\mu$m 
flux density from becoming too bright.  Even an L type star 
produces too much flux at 3.6 $\mu$m at the distance required 
to match the flux density of the WD circumstellar disk (which is 
limited by the size of the WD Roche lobe) to the observed IRAC 
SED.  A model with a T3.5 secondary star and a WD circumstellar 
disk with inner temperature of 1300 K produces an excellent fit 
to the IRAC SED, but only if the distance to the system is a 
mere 35 pc.
This distance is a factor of $\approx4$ smaller than the minimum 
distance quoted in the literature (which was calculated by 
\citealt{araujo05} from fitting the UV spectrum of the WD in 
MR Ser), not to mention requiring a much later spectral type 
for the secondary star than estimated in the literature.  
Consequently, we caution that, although this 
circumstellar-disk-dominated model is suggestive, inasmuch as it 
reproduces the observed IRAC SED so exactly, it is overly complex 
for the ability of our data to constrain it and is highly speculative.

\clearpage


\begin{deluxetable}{lccc}
\tablecaption{Log of observations with the Spitzer Space Telescope\label{t:obslog}}
\tablehead{
\colhead{Polar} & \colhead{AOR Key} & \colhead{Observation Date} & \colhead{Total Exposure Time} \\
\colhead{}      & \colhead{}        & \colhead{(UT)}             & \colhead{(s)}
}
\tablewidth{0pt}
\startdata
V834 Cen & 10191360 & 16 Jul 2005  & \phn20 \\
EF Eri   & 10185984 & 17 Jan 2005  & 300 \\
GG Leo   & 13532160 & 06 May 2005  & 120  \\
V347 Pav & 13531648 & 19 Aug 2005  & 300 \\
VV Pup   & 10191872 & 26 Nov 2004  & 120 \\
MR Ser   & 10188288 & 27 Mar 2005  & \phn20 \\
\enddata
\end{deluxetable}

\begin{deluxetable}{cccccccc}
\tablecaption{Measured 2MASS and IRAC flux densities (mJy)\label{t:fluxes}}
\tablewidth{0pt}
\tablehead{
\colhead{Polar} & 
\colhead{J} & 
\colhead{H} & 
\colhead{K$_{\rm s}$} & 
\colhead{IRAC-1} & 
\colhead{IRAC-2} & 
\colhead{IRAC-3} & 
\colhead{IRAC-4} \\
\colhead{} & 
\colhead{(1.2~$\mu$m)} &  
\colhead{(1.7~$\mu$m)} & 
\colhead{(2.2~$\mu$m)} & 
\colhead{(3.6~$\mu$m)} & 
\colhead{(4.5~$\mu$m)} & 
\colhead{(5.8~$\mu$m)} & 
\colhead{(8.0~$\mu$m)}
}
\startdata
V834 Cen & 6.54$^{+0.16}_{-0.16}$ & 5.60$^{+0.11}_{-0.11}$ & 5.22$^{+0.13}_{-0.12}$ & 2.39$^{+0.05}_{-0.05}$ & 2.33$^{+0.06}_{-0.06}$ & 2.04$^{+0.16}_{-0.16}$ & 2.48$^{+0.15}_{-0.15}$ \\[4pt]
EF Eri & 0.21$^{+0.05}_{-0.04}$ & 0.55$^{+0.08}_{-0.07}$ & 0.47$^{+0.08}_{-0.07}$ & 0.71$^{+0.02}_{-0.02}$ & 0.72$^{+0.02}_{-0.02}$ & 0.66$^{+0.07}_{-0.07}$& 0.70$^{+0.05}_{-0.05}$\\[4pt]
GG Leo & 1.57$^{+0.07}_{-0.06}$ & 1.27$^{+0.09}_{-0.08}$ & 0.93$^{+0.09}_{-0.08}$ & 0.25$^{+0.07}_{-0.07}$ & 0.17$^{+0.06}_{-0.06}$ & 0.19$^{+0.03}_{-0.03}$ & 0.17$^{+0.03}_{-0.03}$ \\[4pt]
V347 Pav & 0.53$^{+0.06}_{-0.05}$ & 0.63$^{+0.09}_{-0.08}$ & 0.68$^{+0.10}_{-0.09}$ & 0.87$^{+0.02}_{-0.02}$ & 0.63$^{+0.02}_{-0.02}$ & 0.60$^{+0.07}_{-0.07}$ & 0.57$^{+0.05}_{-0.05}$ \\[4pt]
VV Pup & 0.96$^{+0.06}_{-0.06}$ & 1.11$^{+0.10}_{-0.09}$ & 1.01$^{+0.09}_{-0.08}$ & 0.67$^{+0.02}_{-0.02}$ & 0.52$^{+0.02}_{-0.02}$ & 0.34$^{+0.05}_{-0.05}$ & 0.26$^{+0.04}_{-0.04}$ \\[4pt]
MR Ser & 3.71$^{+0.09}_{-0.09}$ & 3.14$^{+0.11}_{-0.11}$ & 3.04$^{+0.12}_{-0.11}$ & 1.30$^{+0.03}_{-0.03}$ & 1.44$^{+0.04}_{-0.04}$ & 1.45$^{+0.12}_{-0.12}$ & 1.17$^{+0.07}_{-0.07}$ \\[4pt]
\enddata
\end{deluxetable}

\begin{deluxetable}{llllllllll}
\tablewidth{0pt}
\tabletypesize{\scriptsize}
\rotate
\tablecaption{Polar system parameters from the literature\label{t:parameters}} 
\tablehead{
\colhead{Polar} & 
\colhead{$P_{\rm orb}$} & 
\colhead{$i$} & 
\colhead{$T_{\rm WD}$} & 
\colhead{Secondary} & 
\colhead{$M_{\rm WD}$} & 
\colhead{$M_{2}$} & 
\colhead{Distance} & 
\multicolumn{2}{c}{B (MG)} \\
\colhead{} & 
\colhead{(minutes)} & 
\colhead{($^{\circ}$)} & 
\colhead{(1000 K)} & 
\colhead{Spec.\ Type} & 
\colhead{($M_{\odot}$)} & 
\colhead{($M_{\odot}$)} & 
\colhead{(pc)} & 
\colhead{Dipole} & 
\colhead{Pole 1, Pole2} 
}
\startdata
V834 Cen & 101.517 [20] & 50(10) [16] & 14.3(0.9) [1] & M7.5--8.5 [10] & 0.66(18) [16] & 0.06 [21] & $144^{+18}_{-23}$ [1] & 23 [4] & \nodata \\[4pt]

EF Eri & 81.022932(8) [11] & 55(5) [11] & 9.5(0.5) [2] & M6+/L4--5 [2,7] & 0.6 [2] & 0.055 [9] & 42--120 [2,7] & 13.8 [9] & 16.5, 21 [5] \\[4pt]

GG Leo & 79.87946(7) [3] & 50--60 [3] & 8--12 [3] & M5+ [3] & 1.13(3) [14] & 0.09 [3] & 100--300 [3] & 20--30 [3] & \nodata \\[4pt]

V347 Pav & 90.082219(52) [14] & 64--72 [12] & $11.8^{+0.6}_{-0.5}$ [1] & \nodata  & 1.00(6) [14] & 0.16 [13] & $171^{+33}_{-38}$ [1] & \nodata & 15, 20 [12] \\[4pt]

VV Pup & 100.436 [20] & 78(5) [20] & $11.9^{+0.6}_{-0.5}$ [1] & M7 [8] & 0.73(5) [8] & 0.10(2) [8] & $129^{+29}_{-21}$ [8] & \nodata & 32,\nodata [18] \\[4pt]

MR Ser & 113.46890(14) [15] & 50(5) [19] & 14.2(0.9) [1] & M5--8 [17,6] & 0.7 [20] & 0.1 [20] & $160^{+18}_{-26}$ [1] & 28 [20] & 24.5,\nodata [20] 
\enddata
\tablerefs{
[1] = \citet{araujo05}, 
[2] = \citet{beuermann00}, 
[3] = \citet{burwitz98}, 
[4] = \citet{ferrario92}, 
[5] = \citet{ferrario96}, 
[6] = \citet{harrison05}, 
[7] = \citet{howell01}, 
[8] = \citet{howell06b}, 
[9] = \citet{howell06c}, 
[10] = \citet{mennickent04}, 
[11] = \citet{piirola87}, 
[12] = \citet{potter00}, 
[13] = \citet{ramsay96}, 
[14] = \citet{ramsay04}, 
[15] = \citet{schwope91}, 
[16] = \citet{schwope93a}, 
[17] = \citet{schwope93}, 
[18] = \citet{wick82}, 
[19] = \citet{wickramasinghe91}, 
[20] = \url{http://cvcat.net} and references therein, 
[21] = calculated from other parameters (see text)
}
\tablecomments{With the exception of distance and secondary star spectral type, the values of these parameters are held constant during our modeling.  When there is a formal uncertainty or range of values listed here, the fixed value used in the models is the nominal or midpoint parameter value, respectively.}  
\end{deluxetable}

\begin{deluxetable}{rlcccccc}
\tablewidth{0pt}
\tabletypesize{\small}
\rotate
\tablecaption{Model Parameters for EF Eri\label{t:eferi_parms}}
\tablehead{
   \colhead{Component}
 & \colhead{Parameter\tablenotemark{a}}
 & \colhead{Model 1}
 & \colhead{Model 2} 
 & \colhead{Model 3} 
 & \colhead{Model 4}
 & \colhead{Model 5\tablenotemark{b}}
 & \colhead{Model 6\tablenotemark{b}}
}
\startdata
 System: & $d$ (pc)                    & 105 & 105 & 105 & 105 & 105 & 105  \\
 SS:     & $T_{2}$ (K)                 & 1700 & 1700 & 1700 & 1700 & 1790 & 1790 \\
         & Spec. Type                  & L5.0 & L5.0 & L5.0 & L5.0 & L3.5 & L3.5  \\
 CYC:    & $B$ (MG)                    & \nodata & 13.8 & 13.8 & 13.8 & \nodata & 13.8 \\
         & $m$                         & \nodata & 2 & 3 & 5 & \nodata & 2 \\
         & Sum over fields?            & \nodata & Y & Y & N & \nodata & Y \\
         & $f_{\nu\rm{,CYC}}/f_{\nu\rm{,SS}}$ (3.6$\mu$m) & \nodata & 1.71 & 3.12 & 0.34 & \nodata & 1.55 \\
         & $f_{\nu\rm{,CYC}}/f_{\nu\rm{,SS}}$ (8.0$\mu$m) & \nodata & 11.01 & 7.04 & 0.18 & \nodata & 9.34 \\
 CBD:    & $r_{\rm in}$ ($R_{\rm WD}$)  & 73   & \nodata & \nodata & \nodata & 73   & \nodata \\
         & $T_{\rm in}$ (K)             & 665  & \nodata & \nodata & \nodata & 665  & \nodata \\
         & $r_{\rm out}$ ($R_{\rm WD}$) & 7800 & \nodata & \nodata & \nodata & 7800 & \nodata \\
         & $T_{\rm out}$ (K)            & 20   & \nodata & \nodata & \nodata & 20   & \nodata
\enddata
\tablenotetext{a}{Fixed model parameter values for this target include:\ $R_{\rm WD}=0.0125R_{\odot}$ and $R_{2}=0.1R_{\odot}$.  See Table \ref{t:parameters} for the values of other parameters not listed here.}
\tablenotetext{b}{These models refer to the scaled 2MASS data -- see text.}
\end{deluxetable}

\begin{deluxetable}{rlcccccc}
\tabletypesize{\small}
\tablewidth{0pt}
\rotate
\tablecaption{Model Parameters for V347 Pav\label{t:v347pav_parms}}
\tablehead{
   \colhead{Component}
 & \colhead{Parameter\tablenotemark{a}}
 & \colhead{Model 1}
 & \colhead{Model 2} 
 & \colhead{Model 3}
 & \colhead{Model 4}
 & \colhead{Model 5\tablenotemark{b}}
 & \colhead{Model 6\tablenotemark{b}}
}
\startdata
 System: & $d$ (pc)                    & 171 & 171 & 171 & 171 & 133 & 133 \\
 SS:     & Spec. Type                  & M6.0 & M6.0 & M6.0 & M6.0 & M6.0 & M6.0 \\
 CYC:    & $B$ (MG)                    & \nodata & 15,20 & 15,20 & 15,20 & \nodata & 15,20 \\
         & $m$        & \nodata & 2 & 3 & 2 & \nodata & 5 \\
         & Sum over fields?            & \nodata & Y & Y & N & \nodata & Y \\
         & $f_{\nu\rm{,CYC}}/f_{\nu\rm{,SS}}$ (3.6$\mu$m) & \nodata & 1.06 & 0.80 & 1.68 & \nodata & 0.62  \\
         & $f_{\nu\rm{,CYC}}/f_{\nu\rm{,SS}}$ (8.0$\mu$m) & \nodata & 2.97 & 1.48 & 0.00 & \nodata & 0.93 \\
 CBD:    & $r_{\rm in}$ ($R_{\rm WD}$)  & 148   & \nodata & \nodata & \nodata & 148 & \nodata \\
         & $T_{\rm in}$ (K)             & 770   & \nodata & \nodata & \nodata & 655 & \nodata \\
         & $r_{\rm out}$ ($R_{\rm WD}$) & 19241 & \nodata & \nodata & \nodata & 15509 & \nodata \\
         & $T_{\rm out}$ (K)            & 20    & \nodata & \nodata & \nodata & 20 & \nodata  
\enddata
\tablenotetext{a}{Fixed model parameter values for this target include:\ $R_{\rm WD}=0.008R_{\odot}$, $R_{2}=0.21R_{\odot}$, and $T_{2}=2850$ K.  See Table \ref{t:parameters} for the values of other parameters not listed here.}
\tablenotetext{b}{These models refer to the scaled 2MASS data -- see text.}
\end{deluxetable}

\begin{deluxetable}{rlccc}
\tabletypesize{\small}
\tablewidth{0pt}
\tablecaption{Model Parameters for VV Pup\label{t:vvpup_parms}}
\tablehead{
   \colhead{Component}
 & \colhead{Parameter\tablenotemark{a}}
 & \colhead{Model 1}
 & \colhead{Model 2} 
}
\startdata
 System: & $d$ (pc)                    & 100 & 129 \\
 SS:     & Spec. Type                  & M7.0 & M7.0  \\
 CYC:    & $B$ (MG)                    & \nodata & 32  \\
         & $m$                         & \nodata & 3   \\
         & Sum over fields?            & \nodata & Y   \\
         & $f_{\nu\rm{,CYC}}/f_{\nu\rm{,SS}}$ (3.6$\mu$m) & \nodata & 1.13  \\
         & $f_{\nu\rm{,CYC}}/f_{\nu\rm{,SS}}$ (8.0$\mu$m) & \nodata & 1.35 \\
 CBD:    & $r_{\rm in}$ ($R_{\rm WD}$)  & 108 & \nodata \\
         & $T_{\rm in}$ (K)             & 510 & \nodata  \\
         & $r_{\rm out}$ ($R_{\rm WD}$) & 8140 & \nodata  \\
         & $T_{\rm out}$ (K)            & 20 & \nodata  
\enddata
\tablenotetext{a}{Fixed model parameter values for this target include:\ $R_{\rm WD}=0.0105R_{\odot}$, $R_{2}=0.15R_{\odot}$, and $T_{2}=2670$ K.  See Table \ref{t:parameters} for the values of other parameters not listed here.}
\end{deluxetable}

\begin{deluxetable}{rlcccc}
\tabletypesize{\small}
\tablewidth{0pt}
\tablecaption{Model Parameters for V834 Cen\label{t:v834cen_parms}}
\tablehead{
   \colhead{Component}
 & \colhead{Parameter\tablenotemark{a}}
 & \colhead{Model 1}
 & \colhead{Model 2} 
 & \colhead{Model 3} 
}
\startdata
 System: & $d$ (pc)                    & 121 & 121 & 121 \\
 SS:     & Spec. Type                  & M8.0 & M8.0 & M8.0 \\
 CYC:    & $B$ (MG)                    & \nodata & 23 & 23 \\
         & $m$        & \nodata & 2 & 6  \\
         & Sum over fields?            & \nodata & Y & Y  \\
         & $f_{\nu\rm{,CYC}}/f_{\nu\rm{,SS}}$ (3.6$\mu$m) & \nodata & 6.57 & 6.66  \\
         & $f_{\nu\rm{,CYC}}/f_{\nu\rm{,SS}}$ (8.0$\mu$m) & \nodata & 11.39 & 8.37  \\
 CBD:    & $r_{\rm in}$ ($R_{\rm WD}$)  & 95 & \nodata & \nodata \\
         & $T_{\rm in}$ (K)             & 875 & \nodata & \nodata  \\
         & $r_{\rm out}$ ($R_{\rm WD}$) & 14663 & \nodata & \nodata  \\
         & $T_{\rm out}$ (K)            & 20 & \nodata & \nodata 
\enddata
\tablenotetext{a}{Fixed model parameter values for this target include:\ $R_{\rm WD}=0.0115R_{\odot}$, $R_{2}=0.10R_{\odot}$, and $T_{2}=2490$ K.  See Table \ref{t:parameters} for the values of other parameters not listed here.}
\end{deluxetable}

\begin{deluxetable}{rlcc}
\tabletypesize{\small}
\tablewidth{0pt}
\tablecaption{Model Parameters for GG Leo\label{t:ggleo_parms}}
\tablehead{
   \colhead{Component}
 & \colhead{Parameter\tablenotemark{a}}
 & \colhead{Model 1}
 & \colhead{Model 2} 
}
\startdata
 System: & $d$ (pc)                    & 200 & 200 \\
 SS:     & Spec. Type                  & M7.5 & M7.5  \\
 CYC:    & $B$ (MG)                    & \nodata & 25  \\
         & $m$        & \nodata & 10   \\
         & Sum over fields?            & \nodata & N   \\
         & $f_{\nu\rm{,CYC}}/f_{\nu\rm{,SS}}$ (3.6$\mu$m) & \nodata & 1.20  \\
         & $f_{\nu\rm{,CYC}}/f_{\nu\rm{,SS}}$ (8.0$\mu$m) & \nodata & 0.00  \\
 CBD:    & $r_{\rm in}$ ($R_{\rm WD}$)  & 185 & \nodata \\
         & $T_{\rm in}$ (K)             & 580 & \nodata  \\
         & $r_{\rm out}$ ($R_{\rm WD}$) & 16504 & \nodata  \\
         & $T_{\rm out}$ (K)            & 20 & \nodata 
\enddata
\tablenotetext{a}{Fixed model parameter values for this target include:\ $R_{\rm WD}=0.006R_{\odot}$, $R_{2}=0.13R_{\odot}$, and $T_{2}=2580$ K.  See Table \ref{t:parameters} for the values of other parameters not listed here.}
\end{deluxetable}

\begin{deluxetable}{rlcc}
\tabletypesize{\small}
\tablewidth{0pt}
\tablecaption{Model Parameters for MR Ser\label{t:mrser_parms}}
\tablehead{
   \colhead{Component}
 & \colhead{Parameter\tablenotemark{a}}
 & \colhead{Model 1}
 & \colhead{Model 2} 
}
\startdata
 System: & $d$ (pc)                    & 134 & 134 \\
 SS:     & Spec. Type                  & M7.0 & M7.0  \\
 CYC:    & $B$ (MG)                    & \nodata & 28  \\
         & $m$                         & \nodata & 2   \\
         & Sum over fields?            & \nodata & Y   \\
         & $f_{\nu\rm{,CYC}}/f_{\nu\rm{,SS}}$ (3.6$\mu$m) & \nodata & 3.50  \\
         & $f_{\nu\rm{,CYC}}/f_{\nu\rm{,SS}}$ (8.0$\mu$m) & \nodata & 4.79  \\
 CBD:    & $r_{\rm in}$ ($R_{\rm WD}$)  & 122 & \nodata \\
         & $T_{\rm in}$ (K)             & 715 & \nodata  \\
         & $r_{\rm out}$ ($R_{\rm WD}$) & 14370 & \nodata  \\
         & $T_{\rm out}$ (K)            & 20 & \nodata 
\enddata
\tablenotetext{a}{Fixed model parameter values for this target include:\ $R_{\rm WD}=0.01R_{\odot}$, $R_{2}=0.15R_{\odot}$, and $T_{2}=2670$ K.  See Table \ref{t:parameters} for the values of other parameters not listed here.}
\end{deluxetable}

\begin{deluxetable}{rlcccccc}
\tabletypesize{\small}
\tablewidth{0pt}
\rotate
\tablecaption{Model Parameters for ``Best'' Models\label{t:best_parms}}
\tablehead{
   \colhead{Component}
 & \colhead{Parameter\tablenotemark{a}}
 & \colhead{EF Eri}
 & \colhead{V347 Pav}
 & \colhead{VV Pup}
 & \colhead{V834 Cen}
 & \colhead{GG Leo}
 & \colhead{MR Ser}
}
\startdata
 System: & $d$ (pc) & 105 & 171 & 129 & 121 & 200 & 134 \\
 SS:     & Spec. Type & L5.0 & M6.0 & M7.0 & M8.0 & M7.5 & M7.0 \\
 CYC:    & $B$ (MG) & 13.8 & 15,20 & 32 & 23 & 25 & 28 \\
         & $m$ & 3 & 2 & 3 & 6 & 5 & 2 \\
         & Sum over fields? & Y & N & Y & Y & Y & Y \\
         & $f_{\nu\rm{,CYC}}/f_{\nu\rm{,SS}}$ (3.6$\mu$m) & 3.07 & 1.20 & 1.12 & 5.23 & 0.94 & 3.11 \\
         & $f_{\nu\rm{,CYC}}/f_{\nu\rm{,SS}}$ (8.0$\mu$m) & 7.06 & 0.00 & 1.33 & 6.56 & 1.19 & 4.25 \\
 CBD:    & $r_{\rm in}$ ($R_{\rm WD}$)  & 73 & 148 & 108 & 95 & 185 & 122 \\
         & $T_{\rm in}$ (K)             & 555 & 755 & 480 & 805 & 550 & 670 \\ 
         & $r_{\rm out}$ ($R_{\rm WD}$) & 6129 & 18743 & 7508 & 13119 & 15376 & 13177 \\
         & $T_{\rm out}$ (K)            & 20 & 20 & 20 & 20 & 20 & 20 
\enddata
\tablenotetext{a}{Parameters not listed here have the values in the corresponding table from Tables \ref{t:eferi_parms}--\ref{t:mrser_parms} and Table \ref{t:parameters}.}
\end{deluxetable}


\clearpage


\clearpage

\begin{figure}
\plotone{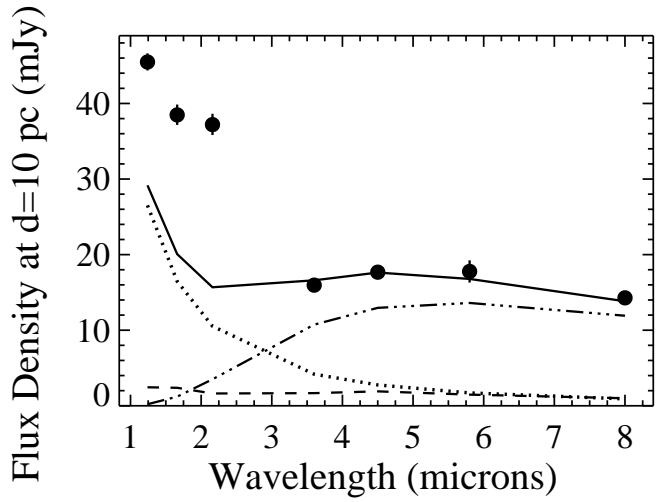}
\caption{Speculative model SED for MR Ser 
showing the observed data (filled circles), total model SED 
(solid line) and model components:\ WD (dotted line), secondary star (short 
dashed line), and WD circumstellar 
disk (dot-dot-dashed line -- see Appendix \ref{s:csd}).  
\label{f:csd_models}}
\end{figure}


\end{document}